\documentclass[cernpreprint,texlive=2016,UKenglish,texmf]{atlasdoc}
 
\usepackage{atlaspackage}
\usepackage{multirow}
\usepackage{feynmp-auto}
\usepackage{siunitx}
\usepackage{booktabs}
\usepackage{atlasbiblatex}
 
\usepackage[process]{atlasphysics}
 
\graphicspath{{logos/}{figures/}}
 
\usepackage{ANA-HIGG-2019-04-PAPER-defs}

\AtlasTitle{Measurements of Higgs Bosons Decaying to Bottom Quarks from Vector Boson Fusion Production with the ATLAS Experiment at $\sqrt{s}=13\,\tev$}
 
\AtlasAbstract{
The paper presents a measurement of the Standard Model Higgs Boson decaying to $b$-quark pairs in the vector boson fusion (VBF) production mode. A sample corresponding to 126~fb$^{-1}$ of $\sqrt{s} = 13\,\tev$ proton-proton collision data, collected with the ATLAS experiment at the Large Hadron Collider, is analyzed utilizing an adversarial neural network for event classification.
The signal strength, defined as the ratio of the measured signal yield to that predicted by the Standard Model for VBF Higgs production, is measured to be  $0.95^{+0.38}_{-0.36}$ , corresponding to an observed (expected) significance of  2.6~(2.8) standard deviations  from the background only hypothesis. The results are additionally combined with an analysis of Higgs bosons decaying to $b$-quarks, produced via VBF in association with a photon.
}

\PreprintIdNumber{CERN-EP-2020-195}
 
\AtlasJournalRef{Eur. Phys. J. C. 81 (2021) 537}
\AtlasDOI{10.1140/epjc/s10052-021-09192-8}
 
\hypersetup{pdftitle={ATLAS document},pdfauthor={The ATLAS Collaboration}}
 
\begin{document}
 
\maketitle

\section{Introduction}
\label{sec:intro}
\begin{figure}[bp]
\begin{center}
\subfloat[]{\includegraphics[width=0.45\textwidth]{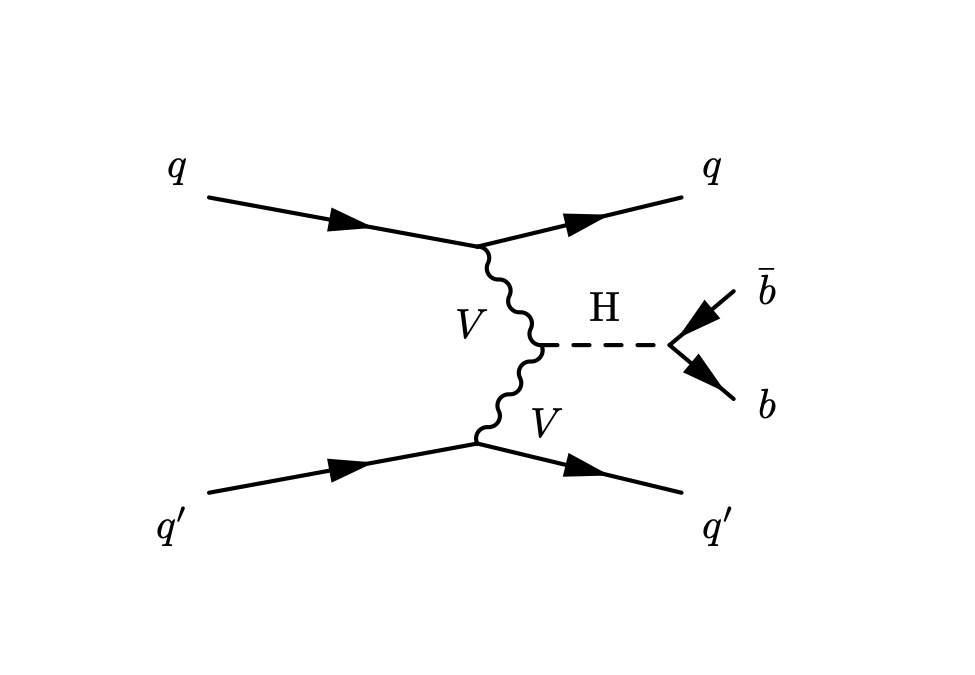}\label{fig:feyn_vbf}}
\subfloat[]{\includegraphics[width=0.45\textwidth]{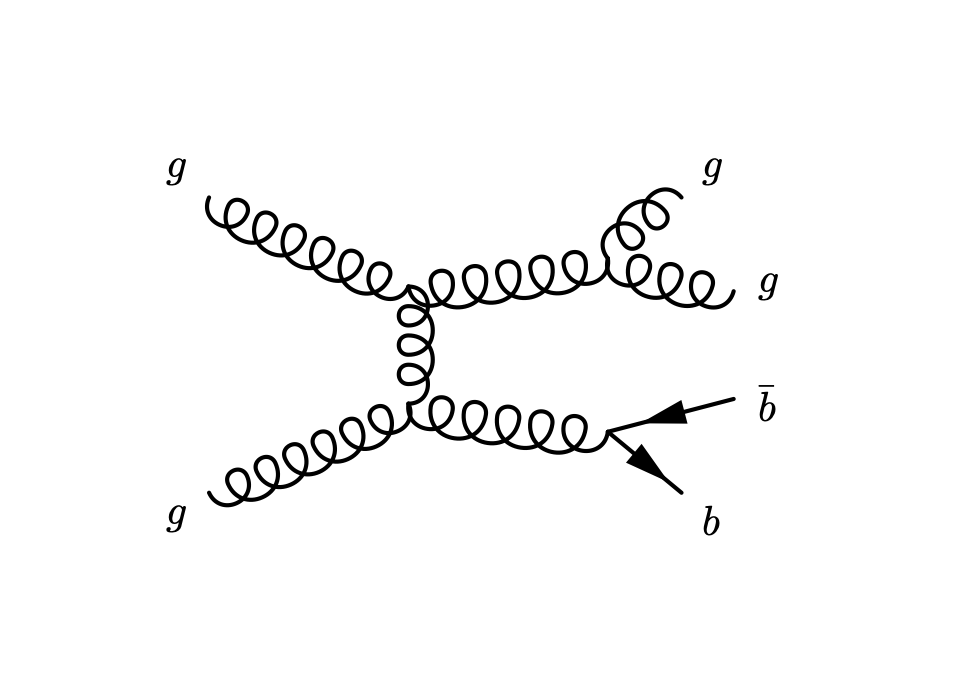}\label{fig:feyn_qcd}}
\caption{Feynman diagrams of (a) the VBF signal process and (b) gluon splitting as an example background process.}
\end{center}
\end{figure}

Approximately 4.5 million Higgs bosons~\cite{Englert:1964et,Higgs:1964ia,Higgs:1964pj,Guralnik:1964eu,Higgs:1966ev,Kibble:1967sv,HIGG-2012-27,CMS-HIG-12-028} have decayed into bottom quarks in the nearly 140~\ifb\ of 13~\TeV\ LHC collisions collected for analysis by the ATLAS experiment during Run 2. While this number is three times larger than that from any other decay mode, this decay channel remains the most poorly measured major Higgs boson decay channel.  In the dominant production mode, where the Higgs boson is produced in a virtual top-quark loop connecting two interacting gluons (ggF), the signature of this decay is overwhelmed by the strong production of quarks and gluons, except in cases where the Higgs boson is highly boosted~\cite{CMS-HIG-19-003}.  The ATLAS and CMS experiments have observed Higgs boson decays into \bquarks, $H\rightarrow b\bar{b}$, with most of the sensitivity coming from cases where the Higgs boson is produced in association with a vector boson (\VH, $V = W,Z$), which provides sufficient discrimination against QCD background processes despite its small cross-section.  The measurement by the ATLAS experiment of the signal yield relative to the Standard Model expectation, $\muH$, is $1.02 \pm 0.12 \text{(stat.)} \pm 0.14 \text{(syst.)}$, corresponding to a significance of $6.7\sigma$~\cite{HIGG-2018-51} relative to the background-only hypothesis. The measurement by the CMS experiment is $\muH= 1.04 \pm 0.14 \text{(stat.)}\pm 0.14 \text{(syst.)}$, corresponding to a significance of $5.6\sigma$~\cite{CMS-HIG-18-016}.
 
Vector-boson fusion (VBF), wherein the Higgs boson is produced when quarks from each proton radiate weak vector bosons that fuse to form the Higgs boson, is the second most frequent Higgs boson production mechanism. 
Its signature, shown in Figure~\ref{fig:feyn_vbf}, is characterized by the presence of jets from each of the quarks with a large rapidity gap between them.  Because there is no coloured connection between the two protons, radiative hadronic activity between the two forward jets is suppressed.  VBF Higgs boson production has been measured by the ATLAS experiment in several decay channels, and the combined result for the signal strength is $\muVBF=1.21 \pm  0.18 \text{(stat.)}\pm 0.15 \text{(syst.)}$~\cite{HIGG-2018-57}. The most recent CMS measurement yields $\muVBF = 0.73 \pm 0.23 \text{(stat.)}\pm 0.16 \text{(syst.)}$ also through the combination of several decay channels~\cite{CMS-HIG-17-031}.
 
Studying Higgs boson decays into \bquarks in the vector-boson fusion channel provides an avenue to pursue this challenging signature. Previous measurements of this process by the ATLAS experiment using $31\,\ifb$ of $\rts = 13\,\tev$ data yielded combined results for VBF production with and without a photon in the final state~\cite{HIGG-2016-30}.  The resulting signal strength of $H\rightarrow b\bar{b}$ was $\muH = 2.5^{+1.4}_{-1.3}$, corresponding to an observed (expected)  significance of $1.9\sigma$~($0.8\sigma$).  The observed signal strength for VBF production of $H\rightarrow b\bar{b}$ when the other Higgs boson production modes are fixed to their Standard Model values, $\muVBFHbb$, was 3.0$^{+1.7}_{-1.6}$, corresponding to an observed (expected) significance of $1.9\sigma$~($0.7\sigma$) relative the background-only hypothesis, which includes non-VBF Higgs boson production modes.

This paper presents an update in the analysis of VBF production without a photon, using 126~\ifb\ of proton--proton ($pp$) collision data collected by the ATLAS experiment at the Large Hadron Collider from 2016 to 2018.  The analysis has significant improvements in sensitivity with respect to the previous version~\cite{HIGG-2016-30}.  A brief outline of this analysis is described in the following. The final state is characterized by two \bjets from the decay of the Higgs boson, as well as two light-quark jets with a large rapidity gap, coming from the outgoing quarks.  In this analysis, two channels corresponding to the available triggers during the data-taking periods and targeting events with and without a high \pT forward jet are used.  After preselection, kinematic properties of the events are used as input to an adversarial neural network (ANN)~\cite{goodfellow2014generative}, trained on signal Monte Carlo (MC) events and data sidebands.   The ANN is constructed such that the classifier output score is independent of the di-\bjet invariant mass.  Each channel is then divided into five regions of varying sensitivity based on the classifier score.  Backgrounds consist of two primary components: non-resonant QCD multi-jet events, an example shown in Figure~\ref{fig:feyn_qcd},  and events containing $Z$ boson decays into \bjets. The $Z\rightarrow b\bar{b}$ contribution is constrained directly from data using an embedding process in which $Z\rightarrow \mu\mu$ events are selected in data, the muons are replaced with \bjets from simulation, and the analysis selections are applied to determine the number of selected $Z\rightarrow b\bar{b}$ events.  The non-resonant background shape is determined from the data and the same shape is used in all regions for each channel.  It is primarily determined from the lowest-score regions.  Potential bias arising from the assumption that the non-resonant background shape is the same in all regions for each channel is determined through a set of shape-bias control regions.  
The VBF Higgs boson signal is extracted through a  simultaneous fit of the signal and background contributions to the di-\bjet invariant mass spectrum in each channel and region.  Lastly, the results are combined with the 132~\ifb\ measurement by the ATLAS experiment in which VBF production of a Higgs boson decaying into \bquarks is accompanied by a photon~\cite{VBFgamma},  henceforth referred to as the photon analysis, for a joint measurement of \muVBF and \muH.
 
 
\section{Detector and Data Samples}
 
\newcommand{\AtlasCoordFootnote}{
ATLAS uses a right-handed coordinate system with its origin at the nominal interaction point (IP)
in the centre of the detector and the \(z\)-axis along the beam pipe.
The \(x\)-axis points from the IP to the centre of the LHC ring,
and the \(y\)-axis points upwards.
Cylindrical coordinates \((r,\phi)\) are used in the transverse plane,
\(\phi\) being the azimuthal angle around the \(z\)-axis.
The pseudorapidity is defined in terms of the polar angle \(\theta\) as \(\eta = -\ln \tan(\theta/2)\).
Angular distance is measured in units of \(\Delta R \equiv \sqrt{(\Delta\eta)^{2} + (\Delta\phi)^{2}}\).}

The ATLAS detector~\cite{PERF-2007-01} at the LHC covers nearly the entire solid angle around the collision point.\footnote{\AtlasCoordFootnote}
It consists of an inner tracking detector surrounded by a thin superconducting solenoid, electromagnetic and hadronic calorimeters,
and a muon spectrometer incorporating three large superconducting toroidal magnets.
 
The inner-detector system is immersed in a \SI{2}{\tesla} axial magnetic field
and provides charged-particle tracking in the range \(|\eta| < 2.5\).
A high-granularity silicon pixel detector covers the vertex region and typically provides four measurements per track,
the first hit normally being in the insertable B-layer installed before Run~2 of the LHC~\cite{ATLAS-TDR-2010-19,PIX-2018-001}.
It is followed by a silicon microstrip tracker, which usually provides eight measurements per track.
These silicon detectors are complemented by the transition radiation tracker (TRT),
which enables radially extended track reconstruction up to \(|\eta| = 2.0\).
The TRT also provides electron identification information
based on the fraction of hits (typically 30 in total) above a higher energy-deposit threshold corresponding to transition radiation.
 
The calorimeter system covers the pseudorapidity range \(|\eta| < 4.9\).
Within the region \(|\eta|< 3.2\), electromagnetic calorimetry is provided by barrel and
endcap high-granularity lead/liquid-argon (LAr) calorimeters,
with an additional thin LAr presampler covering \(|\eta| < 1.8\)
to correct for energy loss in material upstream of the calorimeters.
Hadronic calorimetry is provided by the steel/scintillator-tile calorimeter,
segmented into three barrel structures within \(|\eta| < 1.7\), and two copper/LAr hadronic endcap calorimeters.
The solid angle coverage is completed with forward copper/LAr and tungsten/LAr calorimeter modules
optimized for electromagnetic and hadronic measurements respectively.
 
The muon spectrometer comprises separate trigger and
high-precision tracking chambers measuring the deflection of muons in a magnetic field generated by the superconducting air-core toroids.
The field integral of the toroids ranges between \num{2.0} and \SI{6.0}{\tesla\metre}
across most of the detector.
A set of precision chambers covers the region \(|\eta| < 2.7\) with three layers of monitored drift tubes,
complemented by cathode-strip chambers in the forward region, where the background is highest.
The muon trigger system covers the range \(|\eta| < 2.4\) with resistive-plate chambers in the barrel, and thin-gap chambers in the endcap regions.
 
Interesting events are selected to be recorded by the first-level trigger system implemented in custom hardware,
followed by selections made by algorithms implemented in software in the high-level trigger~\cite{TRIG-2016-01}.
The first-level trigger accepts events from the \SI{40}{\MHz} bunch crossings at a rate below \SI{100}{\kHz},
which the high-level trigger reduces in order to record events to disk at about \SI{1}{\kHz}.
 
The data used in this analysis were collected in 2016, 2017 and 2018~\cite{DAPR-2018-01}.  After applying event-cleaning requirements which ensure that the data are of good quality for triggering and reconstructing jets containing \bquarks, these datasets correspond to integrated luminosities of 24.6~\ifb, 43.6~\ifb, and 57.7~\ifb, respectively, for a total integrated luminosity of 125.9~\ifb.  During the 2016 data-taking, roughly one-fifth of the data taken by ATLAS which was good for physics was affected by an inefficiency in the vertex reconstruction in the high-level trigger, which reduced the efficiency of the algorithms used to identify jets originating from $b$-hadron decays; those events were not retained for further analysis and are not included in the 24.6~\ifb.


\section{Monte Carlo Samples}
\label{sec:sigbkgd}
Simulated events with a Higgs boson mass of 125~\GeV{} are used for the signal modelling. 
The signal models include all major Higgs boson production modes: VBF and ggF, as well as \VH and associated production with a pair of top quarks (\ttH).  Simulated VBF signal events were generated at next-to-leading order (NLO) in QCD with \POWHEGBOXV{v2}~\cite{Nason:2004rx,Frixione:2007vw,Alioli:2010xd,Nason:2009ai}, using the NNPDF3.0~\cite{Ball:2012cx} parton distribution functions (PDFs).  NLO electroweak corrections are calculated using \MGMCatNLO v3.0.1~\cite{Alwall:2014hca} and applied as a function of the generated Higgs boson transverse momentum (\pT).  Simulated ggF samples were generated at next-to-next-to-leading order (NNLO) in QCD with \POWHEGBOXV{v2}~\cite{Hamilton:2013fea,Campbell:2012am,Hamilton:2015nsa} using the NNPDF3.0 PDFs.  Both the VBF and ggF samples use  \PYTHIA 8.212~\cite{Sjostrand:2014zea} for parton showering and fragmentation with the A3 tuned parameter set and the NNPDF2.3LO PDF set~\cite{ATL-PHYS-PUB-2016-017}.   For the VBF sample, the \PYTHIA dipole-recoil scheme~\cite{DIPOLE} is used to improve the modelling of radiation in the central region. \POWHEGBOXV{v2} interfaced to NNPDF3.0 was used to generate events that were showered with \PYTHIA 8.212 to model contributions from \VH~\cite{Alioli_2010,Cullen_2012,Hamilton_2012,Luisoni_2013} and \ttH~\cite{Frixione_2007}. All samples were interfaced to \textsc{EvtGen} v1.6.0~\cite{evtgen} for heavy-hadron decays. 
The $Z$ and non-resonant backgrounds are derived from the data. 
 
Multiple $pp$ collisions were simulated with the soft QCD processes of \PYTHIA 8.212  using the A3 tuned parameter set and the NNPDF2.3LO PDFs.  These additional interactions are overlaid on the hard-scatter interaction of the signal and background samples according to the luminosity profile of the recorded data to model contributions from $pp$ interactions in both the same bunch crossing and neighbouring bunch crossings (pile-up). The response of the ATLAS detector to the generated events is then modelled with a detailed ATLAS detector simulation software~\cite{SOFT-2010-01} based on $\GEANT4$~\cite{Agostinelli:2002hh}.

 
\section{Event Reconstruction and Selection}
\subsection{Event reconstruction}
This analysis uses an all-hadronic final state, and therefore the primary objects of interest are jets.  Jets are reconstructed by applying the \antikt jet clustering algorithm~\cite{Cacciari:2008gp,FastJet} with a radius parameter of $R=0.4$ to particle flow~\cite{PERF-2015-09} objects which are constructed from calorimeter energy clusters and tracks. 
In order to remove jets originating from additional $pp$ interactions, the likelihood-based `jet vertex tagger'~\cite{ATLAS-CONF-2014-018} algorithm is applied to all jets with $\pT <60\,\GeV$ and $|\eta|< 2.5$.  Quality and cleaning cuts are applied to ensure well-measured jets. In addition, $\eta$- and \pT-dependent scale factors are applied to correct the jet energy to the hadronic level~\cite{JESJER}, and a pile-up subtraction algorithm is applied to reduce effects of pile-up contributions to the jet energy.
 
A multivariate \btagging algorithm, MV2c10~\cite{FTAG-2018-01}, is used to identify jets which contain $b$-hadrons.  This algorithm is a boosted decision tree which combines several features of impact parameter distributions of associated tracks with properties of secondary vertices, as well as jet $\pT$ and $\eta$. 
A similar algorithm is used by the trigger, differing only in the implementation of the jet, track and primary vertex definitions used, and, in 2016 data-taking, by the flavour-tagging discriminant.
Two different identification working points of the algorithm are used, corresponding to 77\% and 85\% efficiency for \bjets as measured in a $\ttbar$ sample. In the trigger system, the 60\%, 70\% and 85\% working points are used.  In order to improve the resolution of the invariant mass of \bjet pairs, additional corrections are applied to the energy of \btagged jets.  These corrections account for semileptonic decays of the $b$-hadron (the muon-in-jet correction) and energy resolution effects specific to \bjets (the PtReco correction). They improve the dijet mass resolution by up to 20\%~\cite{HIGG-2016-29}.
 
Photons, electrons and muons are also identified in order to veto events which overlap with other $H \rightarrow b\bar{b}$ analysis channels~\cite{VBFgamma,VHtemp}.  Muons are additionally used in the embedding procedure described in Section~\ref{sec:zbkgd}.  Photons are reconstructed from energy clusters in the electromagnetic calorimeter and calibrated to account for energy losses upstream of the calorimeter as well as leakage outside of the clusters~\cite{EGAM-2018-01}.  Photons which convert in the detector material are reconstructed separately, utilizing tracks to find the conversion vertex.  A tight, cut-based selection is applied, providing good rejection of hadronic jets where a neutral meson carries most of the jet energy.  To further suppress the jet backgrounds, the photons are required to be isolated, meaning that after corrections for pile-up the sum of the transverse energy in a cone of $\Delta R = 0.4$ around the photon candidate must not exceed a threshold which depends on the photon's \pT.  Electrons are also reconstructed utilizing energy clusters in the electromagnetic calorimeter~\cite{EGAM-2018-01}.  Electron candidates must satisfy a loose likelihood-based electron identification criterion, including a loose isolation requirement.  Muons~\cite{PERF-2015-10} are reconstructed using inner detector, calorimeter and muon spectrometer information. In this analysis they satisfy loose identification requirements, including a loose isolation requirement.  Of particular importance for this analysis is a requirement that both electrons and muons are prompt, i.e.\ their impact parameter significance is required to be small. 
An overlap removal procedure is applied  in order to ensure that jets, electrons, muons and photons are not double counted.  Events with identified photons with transverse energy~$(\ET)>15\,\GeV$, or electrons or muons with $\pT>7\,\GeV$ after overlap removal are vetoed. The objects have identical selection requirements to Refs~\cite{VBFgamma,VHtemp} in order to ensure orthogonality.
\subsection{Event selection}
 
There are two channels in this analysis, the \twocentral{} and \threecentral channels, which are distinguished by the presence or absence of a high-\pT forward ($3.2 < |\eta| < 4.5)$ jet in the event.  
The specific selections are noted below and summarized in Table~\ref{tab:evtsel}. In the following, $b_1$ and $b_2$ refer to the \btagged jets forming the highest-\pT \btagged jet pair in the event, and $j_1$ and $j_2$ are two additional jets referred to as VBF jets.  Generally, $j_1$  refers to the highest-$\pT$ and most forward VBF jet, and $j_2$ refers to the second-highest-\pT VBF jet.  Additional requirements on the jets are channel specific and detailed below.
 
The trigger for the \twocentral channel targets events with at least two \btagged jets, one with trigger level $\pT > 80$~\GeV\ and a second with $\pT > 60$~\GeV, and at least one forward jet with $\pT > 45$~\GeV.  The corresponding event selection requires at least one \btagged jet passing the 77\% \btagging working point to have $\pT > 85$~\GeV\ and $|\eta| < 2.5$ ($b_1$), and a second \btagged jet passing the 85\% \btagging working point to have $\pT > 65$~\GeV\ and $|\eta| < 2.5$ ($b_2$). One forward jet is required to have $\pT > 60$~\GeV\ ($j_1$) and an additional jet is required with $\pT > 30$~\GeV\ and $|\eta|<4.5$ ($j_2$).  Reconstructed jets in the event are required to be matched to trigger jets.
 
The \threecentral channel trigger targets events with at least three central jets, two of which are $b$-tagged. The trigger jet thresholds varied over the run period.  The corresponding  event selection requires at least two \btagged jets passing the 77\% \btagging working point to have $\pT > 65$~\GeV\ and $|\eta| < 2.5$ ($b_1$, $b_2$).  Two additional jets are required, one jet having $\pT > 160$~\GeV\ and $|\eta|<3.1$ ($j_1$) and a second requiring $\pT > 30$~\GeV\ and $|\eta|<4.5$ ($j_2$). Furthermore, events are rejected if they have any jets with $\pT > 60$~\GeV\ and $3.2 < |\eta|<4.5$ to maintain orthogonality with the \twocentral{} channel. As in the \twocentral channel, reconstructed jets in the event are required to be matched to trigger jets.

In cases where there is ambiguity when identifying the two VBF jets (e.g.\ when more than two jets not selected as the signal \btagged jets satisfy the $\pT$ and $\eta$ requirements), the pair giving the highest invariant mass of the dijet system, \mJJ, is chosen.
 
To remove a shaping of the invariant mass of the \btagged dijet system, $\mbb$, by the high leading-jet thresholds, a cut on the transverse momentum of the \btagged dijet system, \pTbb, is required.   Both channels require $\pTbb> 150$~\GeV. The \threecentral{} channel additionally requires \mJJ to be greater than 800~\GeV.   The efficiency of the combined selection of both channels for simulated VBF signal events with true Higgs boson rapidity $|y_H| < 2.5$ is 0.8\%, with roughly half of the acceptance coming from each channel.  With respect to simulated VBF signal events with $|y_H| < 2.5$ and $\pT^{H} >200$~\GeV, this selection is 10\% efficient, with roughly 4\% and 6\% contributed from the \twocentral{} and \threecentral{} channels, respectively.

\begin{table}[htbp]
\def\arraystretch{1.4}
\centering
\caption{Event selection criteria for the two channels used in this analysis.
The \pT\ and $|\eta|$ requirements on the jets are used to match trigger selections and flavour-tagging requirements.}
\vspace{0.2cm}
{\footnotesize
\begin{tabular}{ l | l | l }
\hline \hline
\multicolumn{3}{c}{\Twocentral{} Channel Event Selection} \\
\hline
\multicolumn{2}{c|}{$b_1$}                         & $\geq1$ \btagged jet at 77\% efficiency working point with $\pT>85$ \GeV{} and $|\eta|<2.5$  \\
\multicolumn{2}{c|}{$b_2$}                         & $\geq1$ \btagged jet at 85\% efficiency working point with $\pT>65$ \GeV{} and $|\eta|<2.5$  \\
\multicolumn{2}{c|}{$j_1$}                         & $\geq1$ jet with $\pT>60$ \GeV{} and $3.2<|\eta|<4.5$  \\
\multicolumn{2}{c|}{$j_2$}                         & $\geq1$ jet with $\pT>30$ \GeV{} and $|\eta|<4.5$  \\
\multicolumn{2}{c|}{}                        & $\pTbb>150$ \GeV{}                         \\
\hline \hline
\multicolumn{3}{c}{\Threecentral{} Channel Event Selection} \\
\hline
\multicolumn{2}{c|}{$b_1$,$b_2$}                         & $\geq2$ \btagged jets at 77\% efficiency working point with $\pT>65$ \GeV{} and $|\eta|<2.5$  \\
\multicolumn{2}{c|}{$j_1$}                         & $\geq1$ jet with $\pT>160$ \GeV{} and $|\eta|<3.1$  \\
\multicolumn{2}{c|}{$j_2$}                         & $\geq1$ jet with $\pT>30$ \GeV{} and $|\eta|<4.5$  \\
\multicolumn{2}{c|}{}                         & no jets with $\pT>60$ \GeV{} and $3.2<|\eta|<4.5$  \\
\multicolumn{2}{c|}{}                        & $\pTbb>150$ \GeV{}, $\mJJ>800$ \GeV{}                  \\
 
\hline \hline
\end{tabular}
}
\label{tab:evtsel}
\end{table}


\section{Adversarial Neural Network for Event Classification}
 
The sensitivity of the analysis is boosted by using an ANN to divide events into regions of varying signal to background composition.  This  type of multivariate classifier can be constructed such that the network is penalized for learning a feature of the dataset~\cite{louppe2016learning,Shimmin_2017}.  Because the Higgs boson signal is extracted from a fit to \mbb, the ANN used in this analysis is penalized for learning the \mbb{} distribution.  Using this construction, the non-resonant background shape is independent of the classifier score and therefore the same for each region, which allows for extra constraints in the shape of the non-resonant background.

Twelve variables are used as inputs to the ANN:
 
\begin{itemize}
\item  \textbf{\mJJ}: the invariant mass of the VBF jet pair. This is typically larger in the signal than in the background.
\item  \textbf{\pTJJ}: the transverse momentum of the VBF jet pair. This is typically larger in the signal than in the background, due to both the harder \pTbb distribution and the lower jet multiplicity in the signal. 
\item  \textbf{\pTbal}: the ratio of the vectorial and scalar sums of the transverse momenta of $b_1,b_2, j_1$ and $j_2$. Generally, in the signal the four jets tend to be balanced, so the quantity is smaller in the signal than in the background.
\item  \textbf{$(p_{{\text T}}^{j_1} - p_{{\text T}}^{ j_2})/(p_{{\text T}}^{ j_1} + p_{{\text T}}^{ j_2})$}: the asymmetry in the VBF jet transverse momenta, which tends to be smaller in the signal than in the background.
\item  \textbf{\dEtabbJJ}: the separation in $\eta$ between the \btagged jet pair and the VBF jet pair. Due to the dominant $g(\rightarrow b\bar{b})g(\rightarrow jj)$ background, where the
gluons can have a large separation in $\eta$, this quantity tends to be larger in the background than in the signal.
\item  \textbf{\dPhibbJJ}: the separation in $\phi$ between the \btagged jet pair and the VBF jet pair. In the signal, the $b\bar{b}$ and $jj$ systems tend to have a larger separation than in the background.
\item  \textbf{$\tan^{-1}(\tan(\Delta\phi(bb)/2)/\tanh(\Delta\eta(bb)/2))$}:  the measure of the relative angle of $\Delta\eta$ and $\Delta\phi$ between the two \btagged jets. This variable takes on higher values in the signal than in the background.
\item  \textbf{$n_{\text{Jets}}$}: the number of jets with $\pT > 20\,\GeV$ and $|\eta|<4.5$. This quantity is, on average, larger in QCD-produced events, such as the background,  than those arising from electroweak production, such as the signal.
\item  \textbf{min $\Delta R (j_{1(2)})$}: the minimum separation in $\Delta R$ between the (sub)leading VBF jet and any jet in the event which is not a part of the \btagged jet pair or VBF jet pair. If there are none of these jets, this quantity takes the default value of 0.
\item  \textbf{$N_{\text{trk}}^{j_{1(2)}}$}: the number of tracks matched to the (sub)leading VBF jet. Signal VBF jets are typically quark-initiated, while background events have a higher gluon composition, which produces higher track-multiplicity jets. This variable is only valid for jets with $\pT>50~\GeV$ and $|\eta|<2.1$. Jets outside this kinematic region take specific values depending on their \pT and $\eta$.  This information is useful as jet quark versus gluon composition varies as a function of \pT{} and $\eta$.
\end{itemize}
 
The ANN consists of a classifier and an adversary.  The classifier's role is to determine if the event is signal- or background-like.  The adversary's role is to determine the value of \mbb{} in terms of a binned \mbb{} distribution.  Then the two are combined such that the overall network discriminates between signal and background but is penalized if the \mbb{} value is learned, i.e.\ if it can accurately determine the \mbb{} bin.
To achieve this, a three-step training procedure is used~\cite{louppe2016learning}. First the classifier is pre-trained with binary cross-entropy loss, while keeping the adversary weights frozen.  Next, the adversary is pre-trained with categorical cross-entropy loss, keeping the classifier weights fixed.  Third, the classifier and adversary are trained together with a combined loss function, $L$.  The combined training proceeds in two sub-steps for each epoch.  First the classifier is trained with $L = L_{\text{classifier}}- \lambda L_{\text{adversary}}$, keeping the adversary weights frozen, and then the adversary is trained with loss function $L = L_{\text{adversary}}$. The configurable parameter $\lambda$ controls to what extent the adversary impacts the overall loss function, i.e.\ how much the network is penalized for learning \mbb. As a post-training step, the classifier scores are scaled to quantiles of the signal MC distribution with the output values ranging from 0 to 1.

One ANN for each channel is trained in Theano through Keras~\cite{theano,keras}.    The Adam optimizer is used with standard parameters from Ref.~\cite{Shimmin_2017}.   Signal events for training data are taken from the VBF MC sample described in Section~\ref{sec:sigbkgd}.  The background events are taken from data sidebands in the range $70~\GeV <\mbb{}<100~\GeV$ (low mass sideband) and $140~\GeV<\mbb{}<200~\GeV$ (high mass sideband).  This range is slightly larger than the mass range used in the signal extraction fit (see Section~\ref{sec:fit}) to avoid edge effects in the \mbb{} spectrum.  Multi-jet MC events are used for determining the network performance and architecture, but data sidebands are used in the final discriminant training due to the fact that the MC sample does not fully capture the data complexity.   For events in the low (high) mass sideband, the adversary \mbb{} bins have a width of 5 (10)~\GeV, leading to six bins per sideband and a roughly equal number of events per bin in each sideband. 
 
The network architecture is optimized for both discrimination of signal from background and decorrelation of the ANN output score with \mbb. Because a reduction in the correlation between \mbb{} and the classifier score is particularly important in the most sensitive signal regions, both the bulk decorrelation and the decorrelation for events with scores in the top 1\% of ANN outputs are used to select the network architecture.  
The network is constructed with $\lambda = 10$ and two epochs of pre-training for both the adversary and classifier, and the adversary training utilizes only the background events. Four-fold cross validation, in which the data is divided into four random subsamples, or folds, and each fold is tested on an ANN trained with the other three folds, is used to verify that there is no overtraining.  This procedure ensures that the data are always categorized with an independently trained ANN. 
Additional uncertainties due to possible bias arising from using data sidebands in the training in which the Higgs boson mass window cannot be directly checked are assessed in Section~\ref{sec:sys-bias}. Figure~\ref{fig:anndist} shows the distributions of the ANN scores for the VBF signal, ggF events and the low and high mass data sidebands. It can be seen that good separation is achieved between the VBF signal and data sidebands.  Additionally it is noted that the ggF events have a similar score distribution to the data sidebands and that the low and high mass sidebands are in good agreement with each other, as shown in the ratio panel.  The variables which contribute most to the ANN's discrimination power are \mJJ, $n_{\text{Jets}}$, $\pTbal$, \dPhibbJJ, and min~$\Delta R (j_{1(2)})$.
 
\begin{figure}[htbp]
\begin{center}
\subfloat[]{\includegraphics[width=0.4\textwidth]{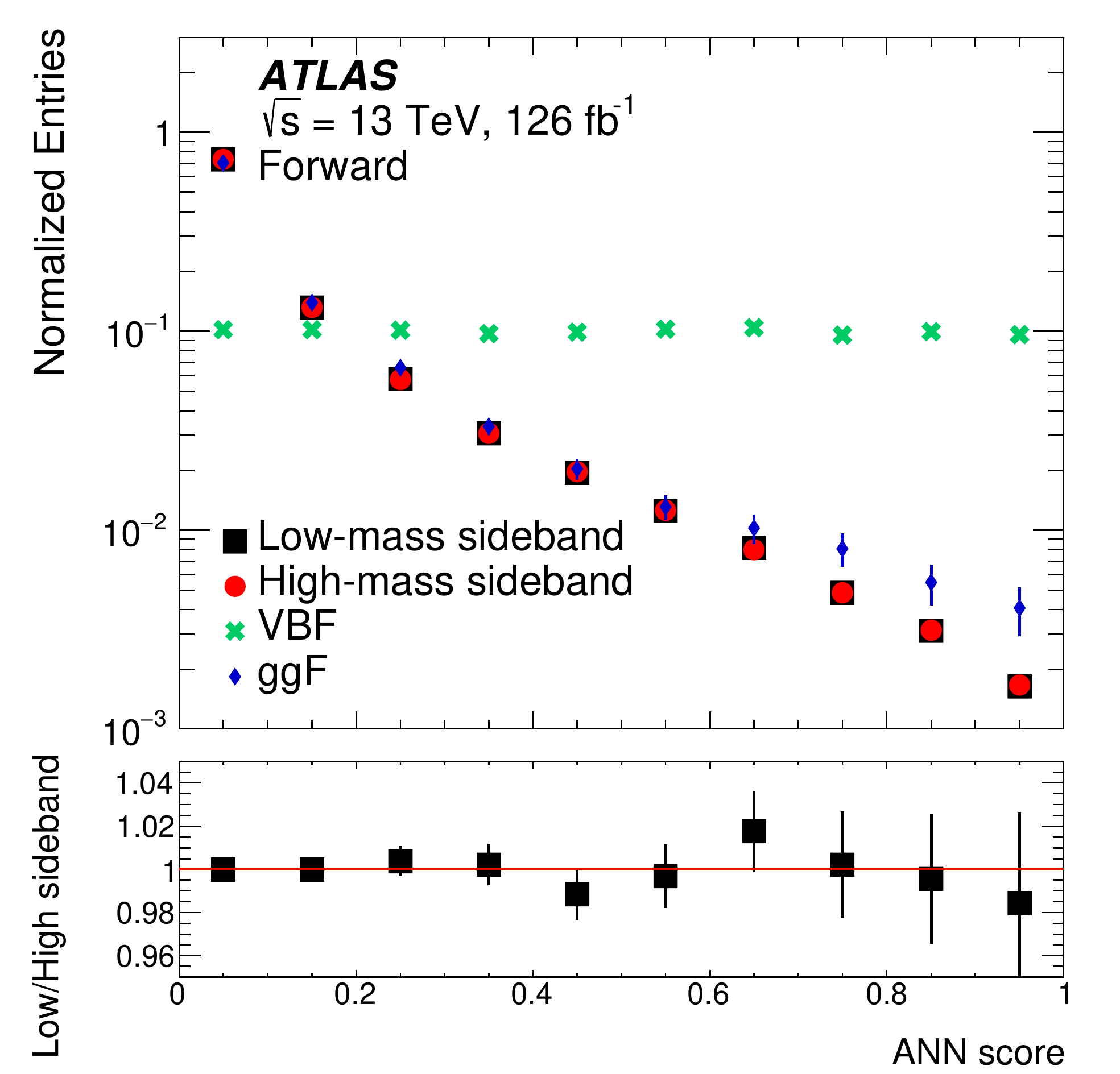}}
\subfloat[]{\includegraphics[width=0.4\textwidth]{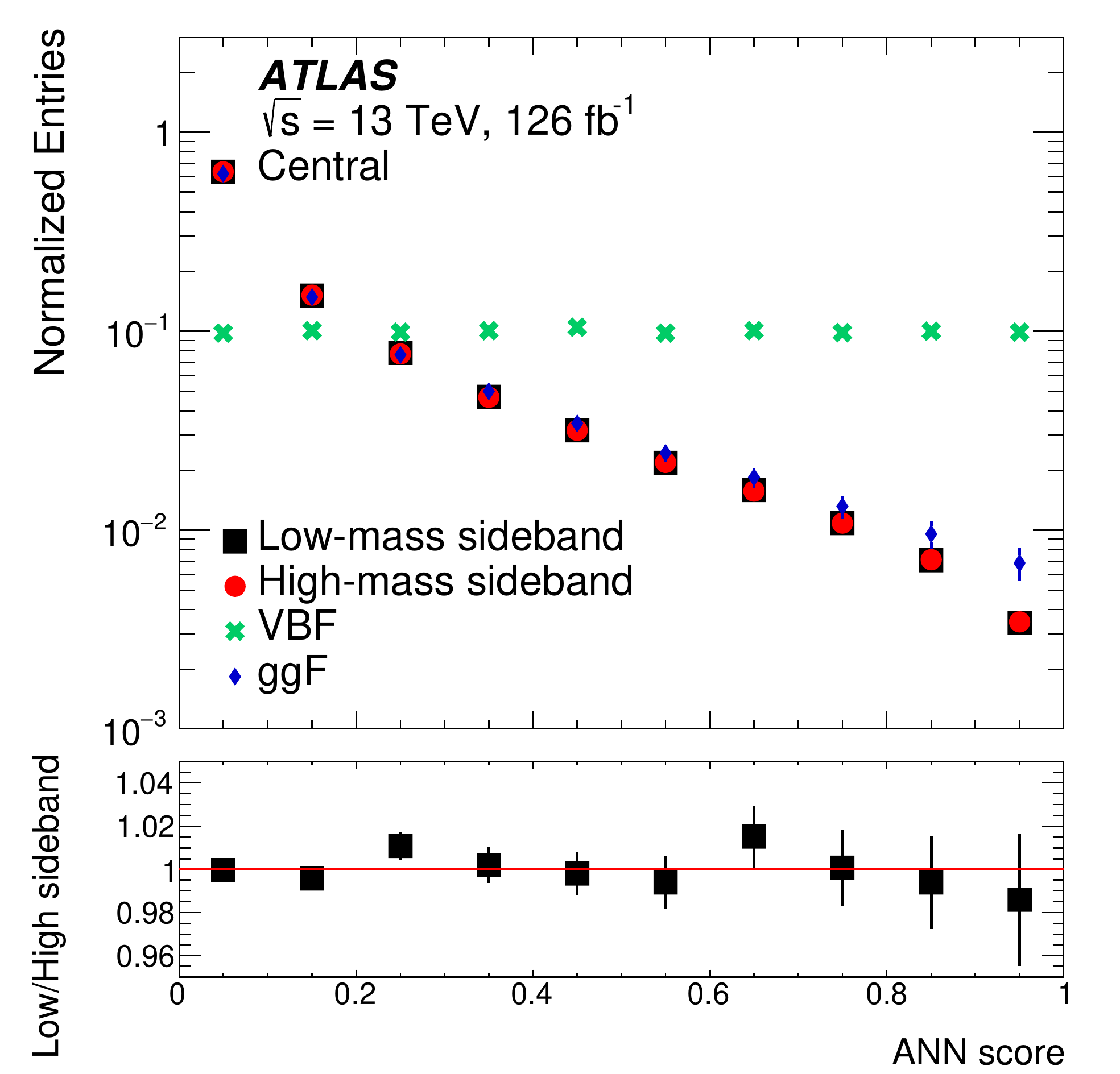}}
\caption{Distribution of ANN output scores for the VBF signal MC events (green crosses), ggF MC events (blue diamonds) and low and high mass data sidebands (black squares, red dots, respectively) for the (a) \twocentral{} channel and (b) \threecentral{} channel. The bottom panel shows the ratio of the low mass to high mass sidebands.  The errors are statistical only.}
\label{fig:anndist}
\end{center}
\end{figure}
 
The ANN score is used to divide the data into signal regions for the signal extraction fit.  Both the number of regions which are used in the fit and the boundaries of those regions are optimized in order to maximize the overall signal sensitivity.   The optimization metric  is
\begin{equation*}
\sum\limits_{n=2}^{N} \frac{s_{n}}{\sqrt{b_{n}}+\delta(b_{n})}
\end{equation*}
where $N$ is the number of regions, $s_{n}$ is the number of signal events in the Higgs boson mass window $(100~\GeV <\, \mbb <\, 140~\GeV)$  in region $n$, $b_{n}$ is the number of background events in the Higgs boson mass window in region $n$, and $\delta(b_{n})$ is the uncertainty on the number of background events.   $s_{n}$ is determined from the signal MC and $b_{n}$ is estimated from a multi-jet MC normalized to the dataside bands.  The uncertainty on the background,  $\delta(b_{n})$, is taken to be 10\% of $b_{n}$.  The significance scan starts from $n=2$ because the lowest classifier score region ($n=1$) is used primarily to constrain the background shape in the high score regions.  
This optimization results in five signal regions for each channel, with region boundaries, signal yields, $Z$ boson yields, and data sideband yields for each channel shown in Table~\ref{tab:regionyield}. The four highest-score regions are named SR1 to SR4; the lowest-score region is named RegLow.
 
\begin{table}
\centering
\caption{For each channel this table shows the lower classifier score boundary of the region (min.\ score), as well as the expected numbers of VBF events, ggF events, and $Z$ events, and the number of events in the combined data sidebands.  Yields for $Z$ event  estimates are determined in data as described in Section~\ref{sec:zbkgd}. Yields from \ttH and \VH events are negligible.  Errors are statistical only.}
\vspace{0.2cm}
{\small
\begin{tabular}{c c c S S[table-format=3.1(2)] S r }
\hline \hline
Channel & Region  & \multicolumn{1}{c}{Min.\ score} & \multicolumn{1}{c}{VBF} &  \multicolumn{1}{c}{ggF}  &  \multicolumn{1}{c}{$Z$}  &  \multicolumn{1}{c}{Data sidebands} \\
\hline
\multirow{5}{*}{\twocentral} & RegLow & 0.00 & 513.3\pm4.7 & 840\pm14 & 13050\pm240 & 2\,524\,065 \\
& SR4 & 0.55 & 126.3\pm2.3 & 12.3\pm1.7 & 158\pm26 & 31\,943 \\
& SR3 & 0.68 & 74.1\pm1.8 & 7.8\pm1.4 & 100\pm21 & 11\,800 \\
& SR2 & 0.76 & 52.8\pm1.5 & 1.6\pm0.6 & 60\pm16 & 6247 \\
& SR1 & 0.82 & 164.4\pm2.6 & 7.8\pm1.4 & 92\pm20 & 10\,649 \\
\hline
\multirow{5}{*}{\threecentral} & RegLow& 0.00 & 530.4\pm4.7 & 923\pm15 & 26040\pm310 & 2\,267\,565 \\
& SR4 & 0.56 & 159.3\pm2.6 & 32.4\pm2.8 & 815\pm55 & 64\,712 \\
& SR3 & 0.73 & 77.4\pm1.8 & 10.0\pm1.6 & 276\pm32 & 19\,490 \\
& SR2 & 0.81 & 45.1\pm1.4 & 4.9\pm1.1 & 174\pm26 & 8838 \\
& SR1 & 0.86 & 134.3\pm2.3 & 8.7\pm1.4 & 183\pm26 & 14\,302 \\
\hline
\end{tabular}
}
\label{tab:regionyield}
\end{table}
 
 
\section{Resonant Background Determination}
\label{sec:zbkgd}
Candidate $Z\rightarrow b\bar{b}$ events contribute significantly to the selected event sample in the range $\mbb <120$~\GeV.  Because of the kinematic requirements of the event selection, simulated $Z$ samples do not accurately reflect the data in the phase space probed by this analysis.  Additionally, the  $Z\rightarrow b\bar{b}$  ANN score distribution is similar to the data sidebands, and consequently there is little constraining power of the resonant contribution in the fit, which is described in Section~\ref{sec:fit}. Therefore, the yield of the resonant component is constrained with a data-driven estimate from $Z\rightarrow \mu^{+}\mu^{-}$ events.  The $Z\rightarrow \mu^{+}\mu^{-}$ candidate data events are selected by requiring a dimuon pair in the $Z$ mass window ($81\,\GeV < m_{\mu \mu} < 101\,\GeV$) recorded by single-muon and double-muon triggers which do not have an isolation requirement.  The reconstructed muon pair in the event is replaced with particle-flow objects obtained by showering, hadronizing and reconstructing an MC-generated \bquark pair with four-vectors matching the muon pair four-vectors.  The \bquark pairs are showered and hadronized with \PYTHIA 8.226 and run through the ATLAS fast simulation~\cite{SOFT-2010-01}, digitization and reconstruction chain.  Jet finding is run on the modified events and the event selection cuts are applied.  The events are first weighted to correct for the muon trigger and offline reconstruction efficiencies. Then the \twocentral{} and \threecentral{} channel trigger requirements are emulated by applying data-driven trigger efficiency maps as a function of the VBF jet \pT{} and $\eta$.   The resulting sample is called the embedded $Z$ sample.  The events are processed through the full event selection and region categorization to yield the number of resonant background events in each region.
 
The process is validated by applying the embedding method in MC events and then comparing embedded and non-embedded events utilizing the analysis selections. Two samples are used for the validation.  Embedded and non-embedded $Z\rightarrow b\bar{b}$ MC events are compared; however, the samples are rather small after analysis selections are applied.  Therefore, the process is also validated using embedded and non-embedded VBF $H\rightarrow b\bar{b}$  events.  The embedded and non-embedded MC samples agree to within 20\% for the variables used in the ANN and for the \mbb{} and \pTbb{} distributions.    In the signal extraction fit described in Section~\ref{sec:fit}, this 20\% residual non-closure is used as an uncertainty in the resonant component.
 
 
\section{Signal Extraction Method}
\label{sec:fit}
The Higgs boson signal strength is determined from an extended binned maximum-likelihood fit to the \mbb{} distribution in data.  The two channels are combined in a joint fit of the \mbb{} distribution where all ten signal regions share the same signal-strength parameter, $\mu$.   The negative log-likelihood is constructed as shown in Equation~\ref{eqn:likelihood},  where $Y_{ijk}$ denotes the yield of $k^{th}$ bin of $j^{th}$ signal region of $i^{th}$ channel. Nuisance parameters, which have external constraints $f(\alpha_l)$, are penalized in the negative log-likelihood.
 
\begin{equation}
\label{eqn:likelihood}
-\log\mathcal{L}(\mu, \{\alpha_{l}\} )= -\log \prod_{i=1}^{\textnormal{Channels}} \prod_{j=1}^{\textnormal{Regions}} \prod_{k=1}^{\textnormal{Bins}} \frac{e^{-Y_{ijk}(\mu, \{\alpha_l\})} \times Y_{ijk}N_{ijk}^{Obs} }{N_{ijk}^{Obs}} 
- \log \prod_{l}^{\textnormal{Nuisance Pars}} f(\alpha_l)
\end{equation}
where
\begin{eqnarray}
Y_{ijk} &=& (\muVBFHbb)N^{\text{VBF}}_{H~ijk}H_{ijk}+ N^{\text{ggF+VH+ttH}}_{H~ijk}H_{ijk} + N_{B~ijk}B_{ik}+ \mu_{z}N_{Z~ijk}Z_{ik} \label{eqn:yield_float} \\
\end{eqnarray}
 
The normalizations of the non-resonant background in each signal region, denoted with $N_B$, are free parameters, whereas the signal normalizations $N^{\text{VBF}}_H$ are set to the Standard Model expectation from simulation.  $\muVBFHbb$ represents the observed signal strength and is an extracted parameter.  The contributions from ggF,  \ttH and \VH production are fixed to their Standard Model predictions, $N^{\text{ggF+VH+ttH}}$.   The measurement is primarily sensitive to differences in the signal contribution relative to the background among the regions, and therefore it is only weakly sensitive to the ggF contribution, which has approximately the same ANN distribution as the backgrounds.  The \ttH and \VH contributions are negligible.  The normalizations of the Z events, $N_Z$, are taken from the embedding process.  The shapes of Higgs boson, Z boson, and non-resonant background are $H$, $Z$ and $B$, respectively.   The Higgs boson signal shape is  determined from a binned fit of a Bukin function~\cite{bukin2007fitting} to MC signal events.  The $Z$ boson shape is also taken from a binned fit of a Bukin function to a sample of $Z\rightarrow b\bar{b}$ events obtained by the embedding process. The normalization of the $Z$ contribution in each region is taken from the number of embedded events passing that region's selection cuts.   For each channel, the non-resonant background $B$ is fit to a binned distribution of arbitrary shape which is subject to the constraint that it is the same in all regions, hence the absence of a $j$ subscript in equations~\ref{eqn:yield_float}. The shape is primarily constrained by each channel's  RegLow region, which has approximately 50 times more events than the other regions in each channel.  The normalization is allowed to float in each region.   The fit range is $80\,\GeV < \mbb{} < 200\,\GeV$.  The bin width is 4~\GeV.  Data in the Higgs boson mass window 
are kept blinded until all the elements of the analysis are finalized.

Systematic uncertainties affecting the signal and simulation-based background components, described in Section~\ref{sec:sys}, are included as nuisance parameters and most are constrained in the likelihood using Gaussian or log-normal probability density functions.  
All experimental uncertainties, described in Section~\ref{sec:sys-exp}, are treated as fully correlated across all regions.   Uncertainties related to different \btagging working points are taken as uncorrelated with each other; however, it was checked that the impact on the sensitivity of treating them as fully correlated or fully uncorrelated is negligible. Theoretical uncertainties, described in Section~\ref{sec:sys-theo}, are treated as correlated across regions. The uncertainty in the resonant background normalization and width related to the embedding process is taken as uncorrelated between regions.   Uncertainties covering possible biases in the non-resonant background shape are included as an additional uncertainty in the signal normalization per region and are uncorrelated between regions.    This bias uncertainty is determined using dedicated control regions described in Section~\ref{sec:sys-bias}.

The parameter of interest in the fit is the signal strength of the vector-boson fusion production channel, \muVBF.   
Tests of Asimov data with signal injected at strengths of $\muVBFHbb = 0,  1, 2$ confirmed the linearity of the fit with no bias in \muVBFHbb.  As a cross-check verifying that the analysis is only weakly sensitive to the overall ggF contribution, a  100\% correlated uncertainty was added to the ggF uncertainties. The expected \muVBFHbb uncertainty changed by only 0.01. The fit is also performed with the inclusive Higgs boson signal strength \muH\ as the parameter of interest.  For this parameter of interest, the ggF, VBF, \ttH and \VH contributions are fixed  to their relative contributions according to the Standard Model values and a single signal strength parameter is applied to all contributions.  Signal injection tests also confirm the linearity of the fit with no bias in \muH.
 
\section{Systematic Uncertainties}
\label{sec:sys}
Systematic uncertainties are divided into three categories: experimental, theoretical and data-driven background uncertainties.  The first two categories impact the signal and simulated component of the resonant background estimation.  All of these uncertainties except the luminosity uncertainty are propagated to the kinematic event variables prior to signal region classification, such that the uncertainty is estimated both for the ANN classification and for the signal and resonant background shape modelling.  The impact of the uncertainties on both the signal region classification and the \mbb{} shape are fully correlated, as are the uncertainties affecting both the signal and the simulated component of the resonant background estimation.
 
\subsection{Experimental systematic uncertainties}
\label{sec:sys-exp}
 
The most prominent experimental uncertainty is due to uncertainties in the \bjet trigger scale factors which account for differences between data and MC events in the efficiency of the triggers.  The per-jet online \btagging efficiency calculated with respect to the offline \btagging efficiency is measured in \ttbar events in both data and MC events.  A scale factor is then applied to the leading simulated \btagged jet  as a function of the leading jet's \pT.  For the 2016 dataset, an additional scale factor binned in the leading jet's $\eta$ was applied.  
The total uncertainty in the \bjet trigger  scale factor is typically 2--5\%.  These uncertainties are applied to signal events and the embedded $Z$ sample which uses simulated \bjets.  Additional jet trigger scale-factor uncertainties, which are determined by taking the difference between the leading jet trigger efficiencies as determined in data and MC events using the same procedure, are approximately 1\%.
 
The second leading experimental uncertainty impacts the per-jet track multiplicity distribution used for quark--gluon separation~\cite{ATL-PHYS-PUB-2017-009} in the ANN.  This quantity is affected by the modelling of the charged-particle multiplicity in jet fragmentation models as well as the track reconstruction efficiency in jets.  The modelling uncertainties are determined through a measurement of the per-jet charged-particle multiplicity in data which is then compared with various models.  The experimental uncertainties are evaluated through standard track-reconstruction efficiency techniques, and their estimation is fully described in Ref.~\cite{ATL-PHYS-PUB-2017-009}.
 
Jet energy scale (JES) and jet energy resolution (JER) uncertainties comprise the next largest group of experimental uncertainties.  The JES uncertainties are primarily determined using data-based $Z$-boson--jet, photon--jet and multi-jet \pT-balancing techniques~\cite{JESJER}. Additional uncertainties are applied for the energy scale of jets containing \bquarks.  The impact of the uncertainty in the JES is estimated by scaling the jet energies within their uncertainties.  JER uncertainties are also determined from \textit{in situ} measurements of $Z$-boson--jet , photon--jet  and dijet \pT balancing~\cite{JESJER}.  The systematic uncertainty due to the JER is calculated by increasing the resolution within its uncertainties, smearing the jet energy by the resulting change in resolution, and comparing the result with the nominal shape and normalization in simulation. 
 
Offline reconstructed \btagging scale factors and their associated uncertainties are determined on a per-jet basis through comparisons of data and MC \ttbar, $W+c$, $D^{*}$, and multi-jet events~\cite{FTAG-2018-01,ATLAS-CONF-2018-001,ATLAS-CONF-2018-006}. For \bjets the uncertainty in the scale factors is approximately 2\%, for \cjets it is 10\%, and for light-quark and gluon jets it is 30\%.
 
The uncertainty in the integrated luminosity is 1.7\%~\cite{ATLAS-CONF-2019-021}, obtained using the LUCID-2 detector~\cite{LUCID2} for the primary luminosity measurements.  This uncertainty is applied to the signal process.  Other uncertainties include the uncertainty from the jet vertex tagging requirement, which is measured on a per-jet basis in $Z\rightarrow \ell^{+}\ell^{-}$ events with~one~jet, and the MC pile-up reweighting uncertainty.
 
\subsection{Theoretical systematic uncertainties}
\label{sec:sys-theo}
 
Theoretical uncertainties primarily affect the signal modelling and are divided into several categories.  The value of the \Hbb branching ratio and its uncertainty are calculated by the HDECAY program~\cite{HDECAY} using the LHC Higgs Cross Section Working Group recommendations~\cite{HiggsWG} with $m_{H} = 125$\,\GeV.  To estimate the effect of missing higher-order terms on the QCD calculations used to predict the VBF and ggF cross-section and acceptance, the chosen renormalization and factorization scales are independently varied by factors of 0.5 and 2.0.   Additional acceptance uncertainties are applied to ggF events in which additional radiation produces a VBF-like topology by having an additional jet pair with an invariant mass greater than 400\,\GeV~\cite{HiggsWG}.   There is a 20\% uncertainty for all events in this category, and additional uncertainties for events with a fifth jet.  The impact of missing electroweak corrections on the signal cross-section and acceptance is calculated using the recommendations from the LHC Higgs Cross Section Working Group, namely that the uncertainty is calculated from the ratio of the cross-section calculated with NLO electroweak corrections using HAWK 2.0 to the leading-order cross-section plus the contributions from photon-induced processes~\cite{HAWK}.  The uncertainty is applied as a function of the Higgs boson \pT.  Uncertainties in the cross-section and acceptance due to the choice of PDF and $\alphas$ are evaluated using the error eigenvectors of the nominal PDF.  Additionally, the overall PDF$+\alphas$ cross-section uncertainty is applied as a uniform uncertainty.  In order to estimate the uncertainty due to the choice of parton shower and underlying-event model, the difference between generator-level signal samples showered with \PYTHIA 8.212 and \HERWIG 7.0~\cite{Bahr:2008pv,Bellm:2015jjp} is determined as a ratio in the variables \mbb{} and $N_{\text{ jets}}$ which is applied to the nominal signal weights. These variables are chosen as they show the largest differences at the truth level.  When calculating the significance and the signal strength, all uncertainties are used.  When calculating the cross-sections, the cross-section uncertainties are removed and only acceptance uncertainties are considered. Contributions from \ttH and \VH are small and a conservative 100\% uncertainty is assigned to their cross-section and acceptance.
 
\subsection{Data-driven background systematic uncertainties}
\label{sec:sys-bias}

Additional uncertainties arise from potential biases in the non-resonant background shape in the Higgs boson mass window due to the fact that the mass window is not used in the ANN training.  Studies of multi-jet MC confirm that training on the sidebands does not induce any bias in the Higgs boson mass window, however these samples do not have sufficient statistics to derive an uncertainty.  Therefore these uncertainties are assessed with shape-bias control regions (CRs). The CRs are constructed by inverting one or more event selection cuts to produce an orthogonal event sample which has similar kinematics to the SR.  Then the CR events undergo a translation and reweighting, described below, to correct the CR kinematics back to the SR kinematics. 
Lastly, the ANN is applied, generating one control region for each region in each channel.
 
The translation is done by smearing the jet $\eta$ of the CR VBF jets with a Gaussian width of $200\,\GeV{} / p_{\text{T,jet}}$.  The new jet $\eta$ is called $\eta^{\prime}$.  The smearing of the original jet $\eta$ value is repeated until the modified event meets the signal region selections.  Table~\ref{tab:evtsel-CR} presents a summary of the event selection in the CRs.  The \twocentral channel CR uses a trigger requiring four central jets, at least two of which are \btagged, and the event selection requires four jets with $|\eta| <$ 2.5, at least two of which are \btagged.   In the \threecentral{} channel CR, the trigger and event selection is the same as in the SR except that the CR requires  $\Delta \phi_{jj} > \pi/4$ and the $\mJJ$ cut is inverted, i.e.\ $\mJJ<\,800$~\GeV.   
While the \mJJ cut is sufficient to ensure orthogonality, the $\Delta \phi_{jj}$ cut improves agreement between CR and SR kinematic distributions.  In order to improve agreement between the SR and CR in the ANN output score distributions, the \twocentral{} channel CR events are sequentially reweighted to match SR kinematics in \nJets, \mJJ, and  \pTJJ, and the \threecentral{} channel CR events are sequentially reweighted in \dEtabbJJ, \nJets, and $\pTbal$.

\begin{table}[htbp]
\caption{Event selection criteria for the \twocentral{} and \threecentral{} channel CRs.
The \pt~ and $|\eta|$ requirements on the offline jets are used to match trigger selections and flavour-tagging requirements. The variables $\eta^{\prime}$ and $\mJJ^{\prime}$ refer to the $\eta$ and \mJJ{} of the VBF jets after the smearing procedure described in the text.}
\vspace{0.2cm}
\label{tab:evtsel-CR}
\def\arraystretch{1.4}
\centering
\footnotesize
\begin{tabular}{ l | l | l }
\hline \hline
\multicolumn{3}{c}{\Twocentral{} Channel CR Event Selection}\\
\hline
\multicolumn{2}{c|}{$b_1$}                         & $\geq1$ \btagged jet at 77\% efficiency working point with $\pT>85$ \GeV{} and $|\eta|<2.5$  \\
\multicolumn{2}{c|}{$b_2$}                         & $\geq1$ \btagged jet at 85\% efficiency working point with $\pT>65$ \GeV{} and $|\eta|<2.5$  \\
\multicolumn{2}{c|}{$j_1$}                         & $\geq1$ jet with $\pT>60$ \GeV{}, $3.2<|\eta^{\prime}|<4.5$, and $|\eta|<2.5$ \\
\multicolumn{2}{c|}{$j_2$}                         & $\geq1$ jet with $\pT>30$ \GeV{}, $|\eta^{\prime}|<4.5$, and $|\eta|<2.5$ \\
\multicolumn{2}{c|}{}                        & $\pTbb>150$ \GeV{}                         \\
\multicolumn{2}{c|}{}                        & no jets with $\pT>60$ \GeV{}, $3.2<|\eta|<4.5$               \\
\multicolumn{2}{c|}{}                        & no jet pair (not including $b_1,b_2$) with $\mJJ>800$ \GeV{}                         \\
\multicolumn{2}{c|}{}                        & $\mJJ^{\prime}>800$ \GeV{}                         \\
\hline
\multicolumn{3}{c}{\Threecentral{} Channel CR Event Selection} \\
\hline
\multicolumn{2}{c|}{$b_1$,$b_2$}                         & $\geq2$ \btagged jets at 77\% efficiency working point with $\pT>65$ \GeV{} and $|\eta|<2.5$  \\
\multicolumn{2}{c|}{$j_1$}                         & $\geq1$ jet with $\pT>160$ \GeV{} and $|\eta|<3.1$  \\
\multicolumn{2}{c|}{$j_2$}                         & $\geq1$ jet with $\pT>30$ \GeV{} and $|\eta|<4.5$  \\
\multicolumn{2}{c|}{}                         & no jets with $\pT>60$ \GeV{} and $3.2<|\eta|<4.5$  \\
\multicolumn{2}{c|}{}                        & $\pTbb>150$ \GeV{}, \mJJ<800 \GeV{}                  \\
\multicolumn{2}{c|}{}                        &     $\Delta \phi_{jj} > \pi/4$ \\
\multicolumn{2}{c|}{}                        & $\mJJ^{\prime}>800$ \GeV{}                         \\
\hline \hline
\end{tabular}
\end{table}

To determine the bias uncertainty, the CR \mbb{} distribution is reweighted to match the SR \mbb{} using a reweighting factor derived from RegLow in each channel.  This reweighting is performed by first doing a simultaneous fit of each CR and its RegLow to the $Z$ and non-resonant background components.  As in the fit described in Section~\ref{sec:fit}, the non-resonant background has arbitrary shape and the $Z$ normalization and shape is determined using constraints from the embedding.  Next, the $Z$ component is subtracted from the CRs such that only the non-resonant background remains.  Then the embedded $Z$ component is subtracted from the SR RegLow, leaving only the SR RegLow non-resonant background, as the contamination from the signal is negligible.  The ratio of the $Z$-subtracted SR and CR RegLow \mbb{} shapes is determined.  This ratio varies smoothly from 0.95 to 1.1 (1.05) in the \twocentral{} (\threecentral{}) channel. A third-order polynomial fits it well in the range $80\,\GeV\,< \mbb{} < 180\,\GeV$, with a $\chi^{2}$ probability of 0.92 (0.19) in the \twocentral{} (\threecentral) region.  The ratio is applied to all of the $Z$-subtracted CRs for each channel.  Figure~\ref{fig:score_srcr} shows the distribution of ANN scores for the SR events, the CR events prior to reweighting, the CR events after kinematic reweighting, and the CR events after kinematic and \mbb{} reweighting.  It can be seen that after kinematic reweighting the SR and CR ANN distributions agree well. 
There is good agreement between the reweighted CR and SR \mbb{} sidebands for all regions, with $\chi^{2}$ probabilities ranging from 0.20 to 0.91.  Lastly, the $Z$ component is added back to each CR, and each CR is fit for a bias signal simultaneously with RegLow for that channel's control region.  The shape of the bias signal is the same as the Higgs boson signal shape.  The mean of the bias signal is scanned from 100 to 140~\GeV{} and the largest bias signal found in this range is taken as an additional uncertainty in the number of Higgs boson events in that region for the Higgs boson signal-strength fit.  True signal contamination in the control regions are negligible.  The maximum observed bias ranges from 60\% of the expected signal in the least sensitive signal regions to 14\% of the expected signal in the most sensitive signal region.  The bias uncertainties are uncorrelated between channels and regions because they are dominated by statistical effects. Assuming the bias uncertainties to be correlated leads to a modest increase in the overall uncertainty and no change in the central value of the fit.  Several modifications to the CR definitions, smearing procedure and reweighting schemes, including using no reweighting, were tested and no significant change in the bias uncertainty was found.

\begin{figure}[t!]\centering
\subfloat[]{\includegraphics[width=0.4\textwidth]{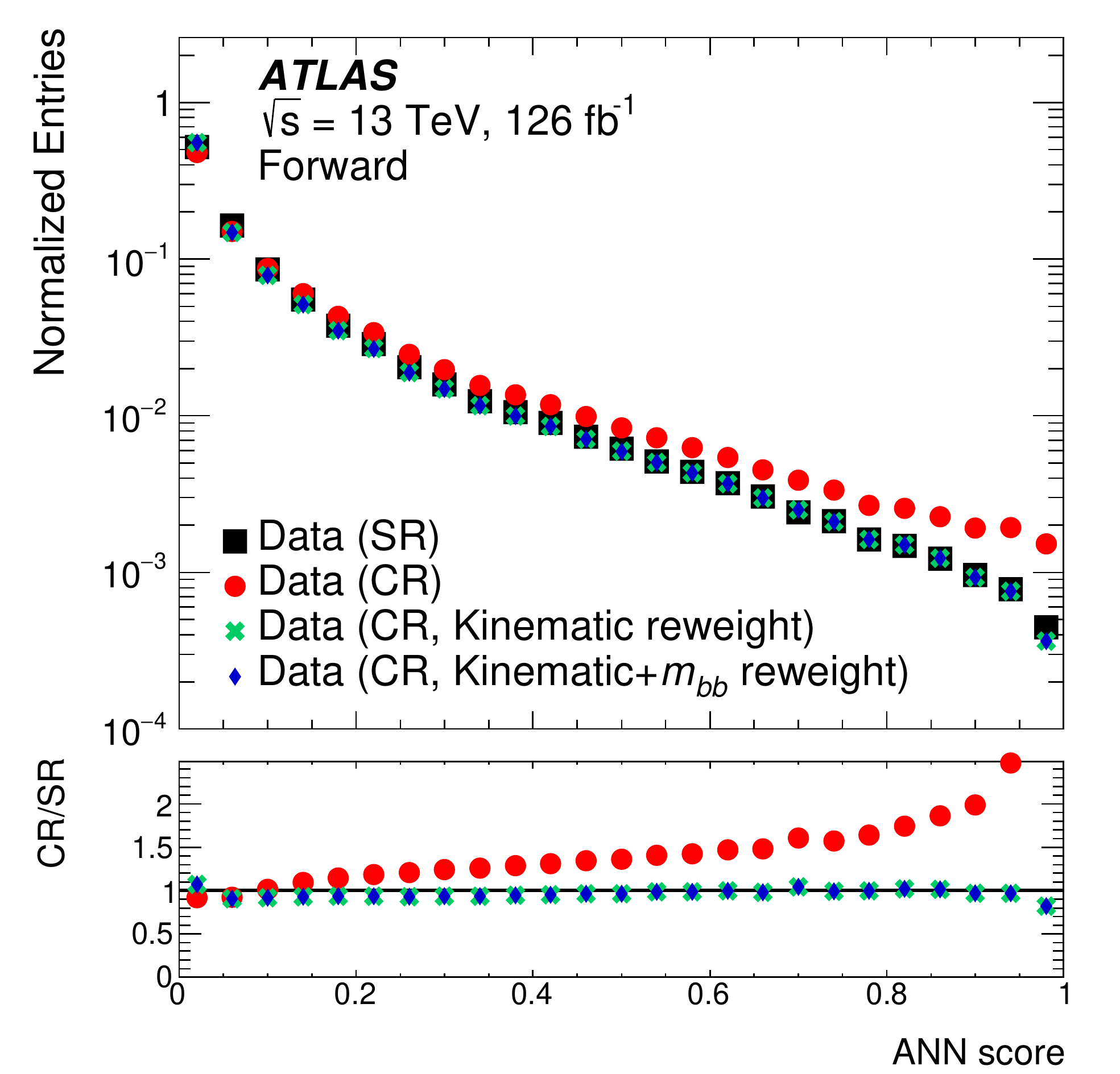}}
\subfloat[]{\includegraphics[width=0.4\textwidth]{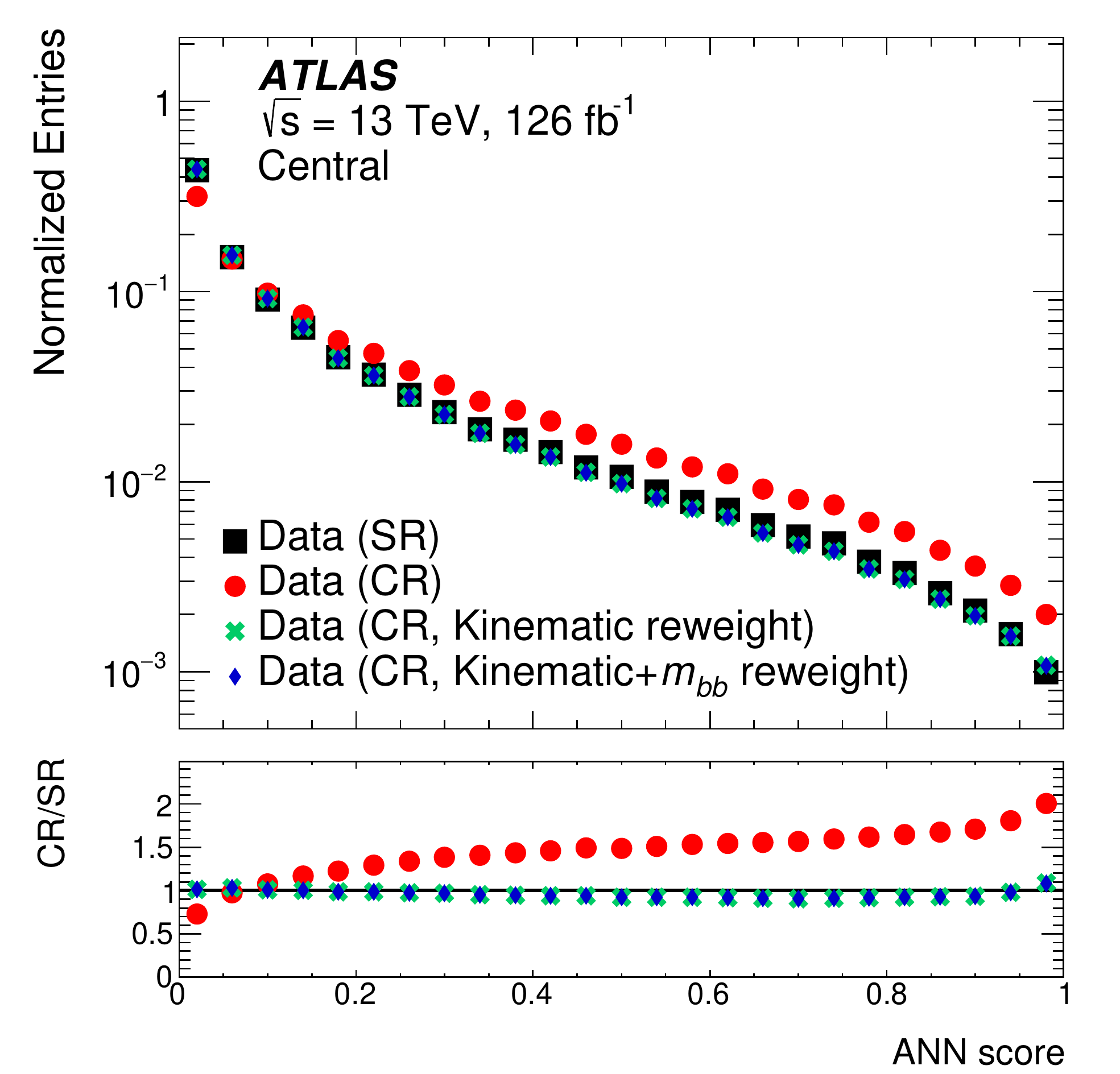}}
\caption{Comparison of the ANN score distributions for the signal and control regions:  (a) shows the \twocentral{} channel, (b) shows the \threecentral{} channel.  The SR selected data are shown in black squares, the CR selected data are shown in red circles prior to reweighting,  in green crosses after kinematic reweighting and in blue diamonds after kinematic and \mbb{} reweighting . The bottom panels show the ratio of the CR distributions to the SR distribution.}
\label{fig:score_srcr}
\end{figure}

The uncertainty in the resonant background normalization related to the embedding process is determined from the 20\% residual non-closure of the method seen in the variables used in the ANN.  Studies show that the uncertainty in the Higgs boson production signal-strength parameter is not sensitive to variations in the resonant background width uncertainty for a large range of non-zero values, and therefore a conservative 20\% is taken as the width uncertainty.  These uncertainties are in addition to the experimental uncertainties related to jet reconstruction, triggering and \btagging described above.

\section{Results}
The fits to the full mass region are shown in Figures~\ref{fig:unblind_1for2cen_pol_full} and~\ref{fig:unblind_3cen_pol_full} for all signal regions in the \twocentral{} and \threecentral{} channels, respectively.  A summary plot, with all signal regions summed, weighted by $\ln(1+s/b)$, is shown in Figure~\ref{fig:unblind_sOverBweight}.  The values of $s$ and $b$ are the integrated numbers of signal and background events in a region centred on the Higgs boson mass and containing 68\% of the signal events.  This weighting approximates the relative contribution each region has in the overall significance of the measurement.  The quality of the fit was checked by performing an eight-parameter-of-interest fit to separately determine the signal strength in each of the four signal regions of each channel.  The probability of compatibility of the signal strengths with a single value is 88\% and the results for the two most sensitive signal regions are consistent, well within one standard deviation of each other.
 
\begin{figure}[t!]\centering
\subfloat[]{\includegraphics[width=0.4\textwidth]{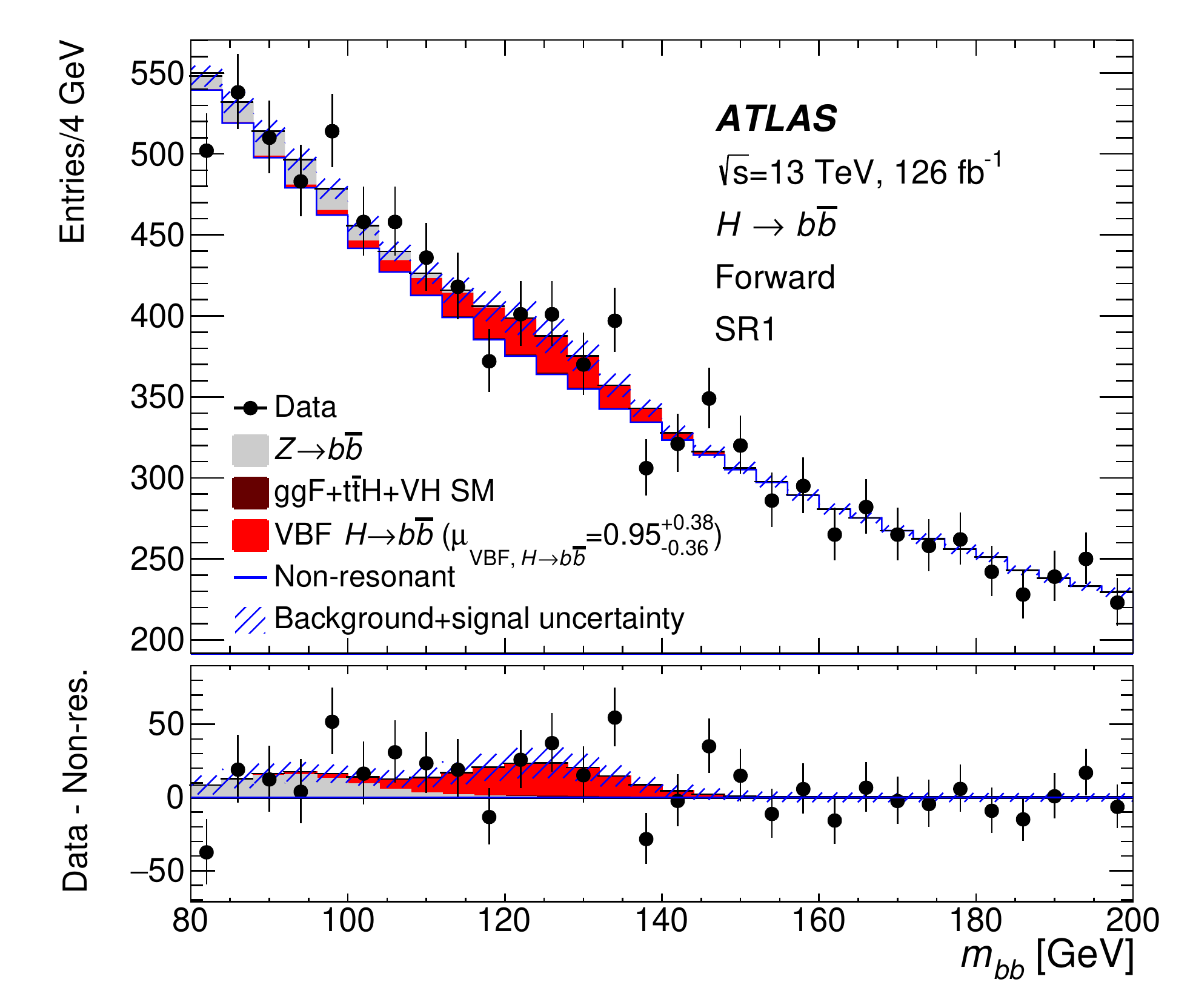}}
\subfloat[]{\includegraphics[width=0.4\textwidth]{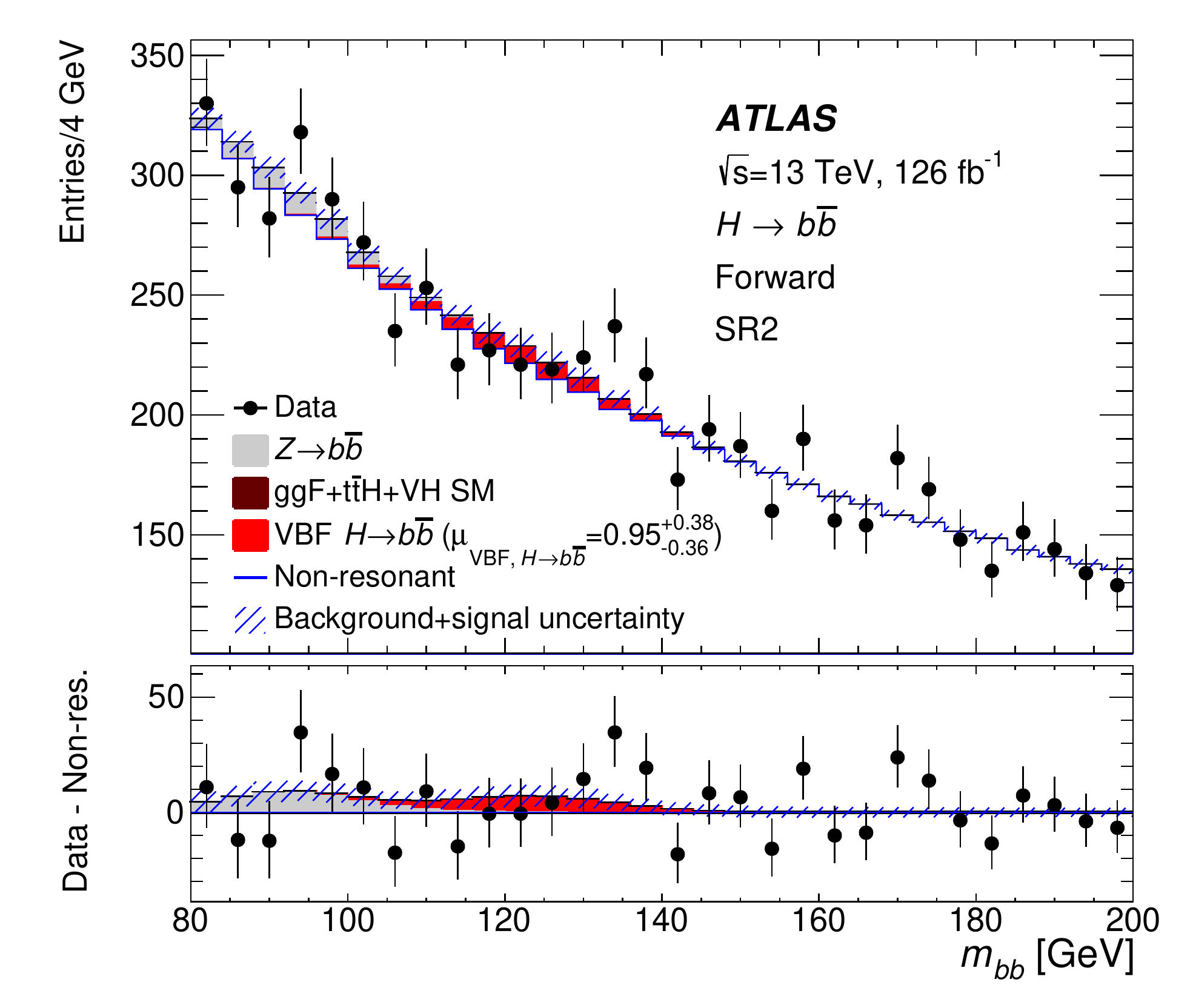}}\\
\subfloat[]{\includegraphics[width=0.4\textwidth]{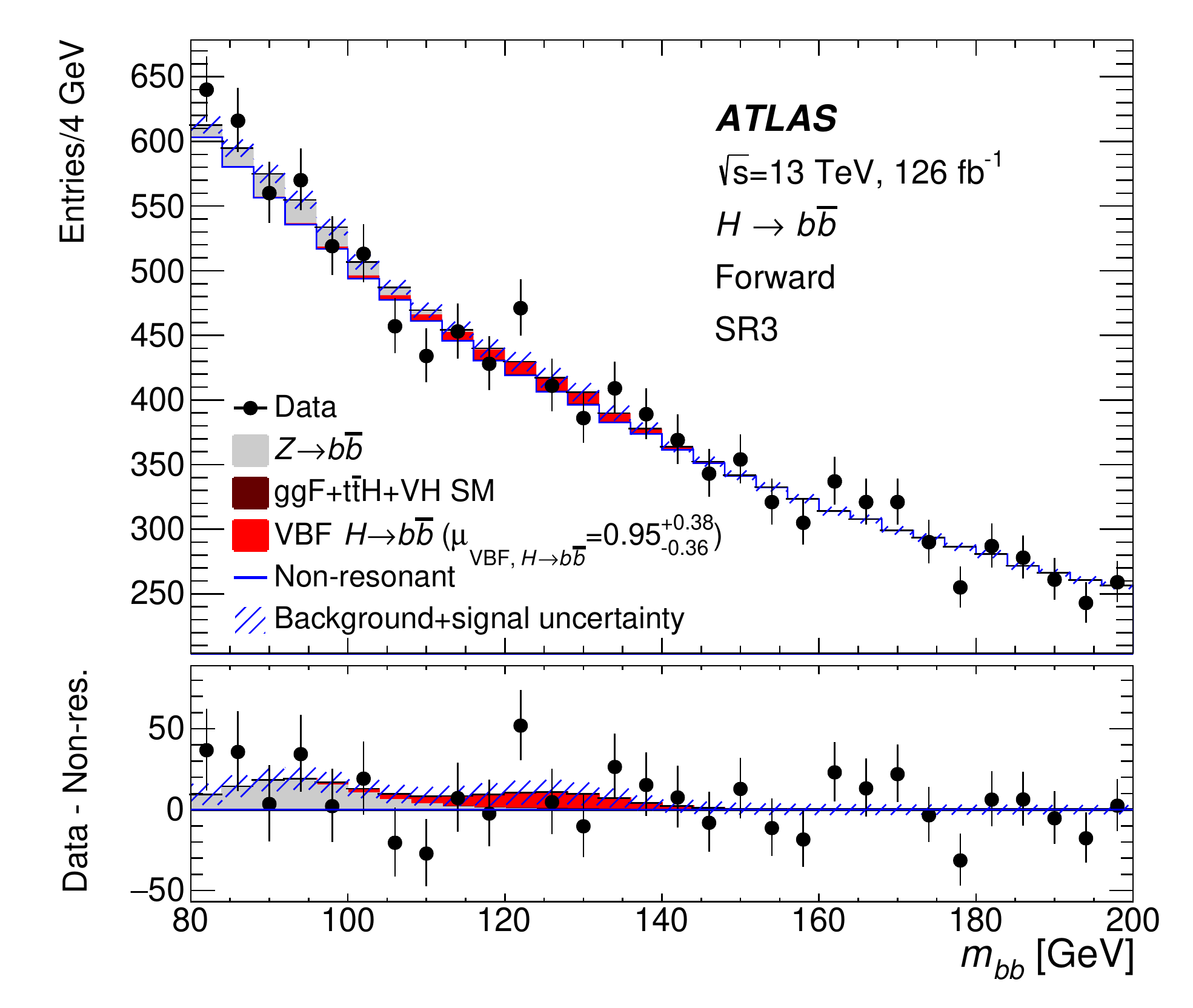}}
\subfloat[]{\includegraphics[width=0.4\textwidth]{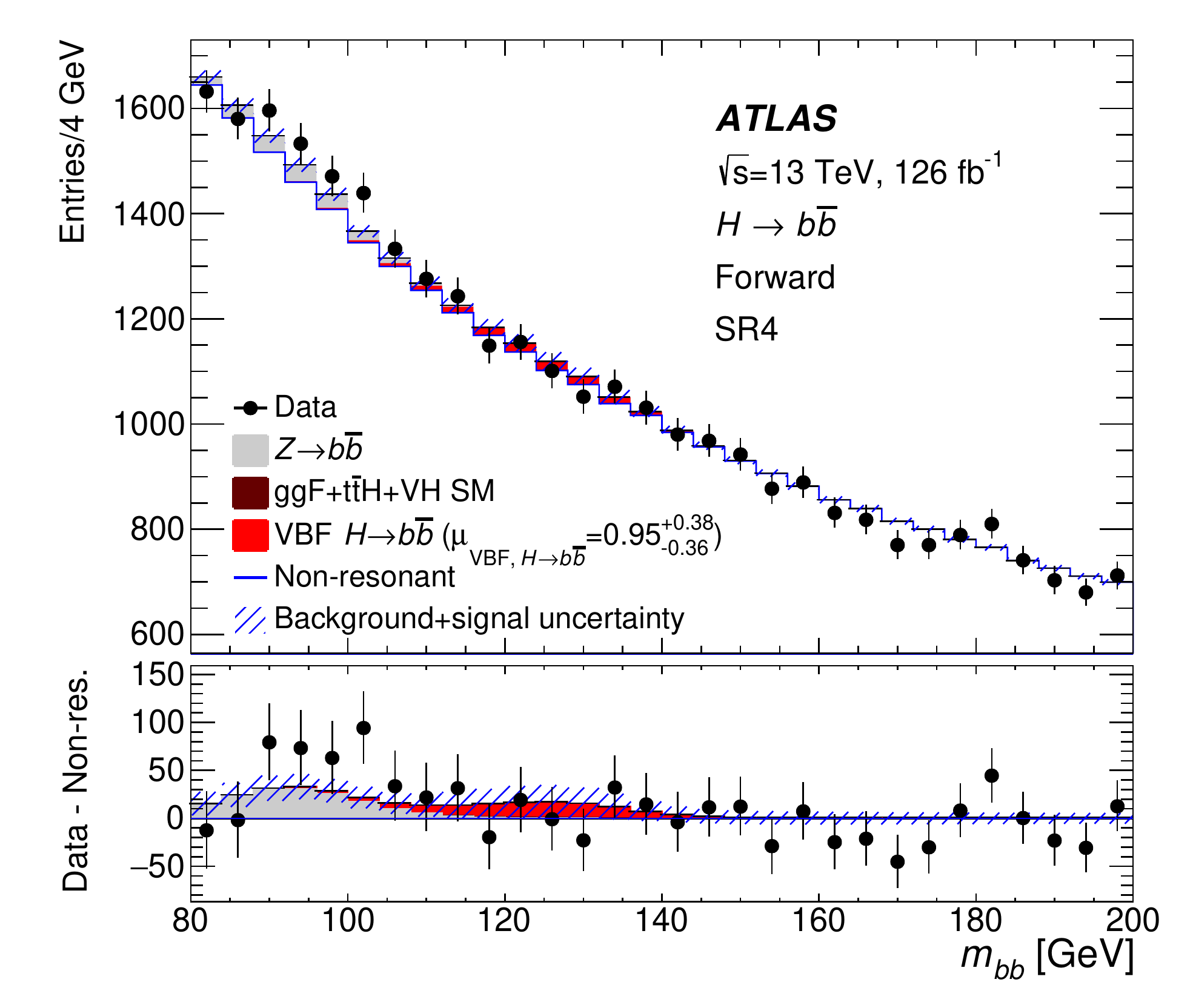}}\\
\subfloat[]{\includegraphics[width=0.4\textwidth]{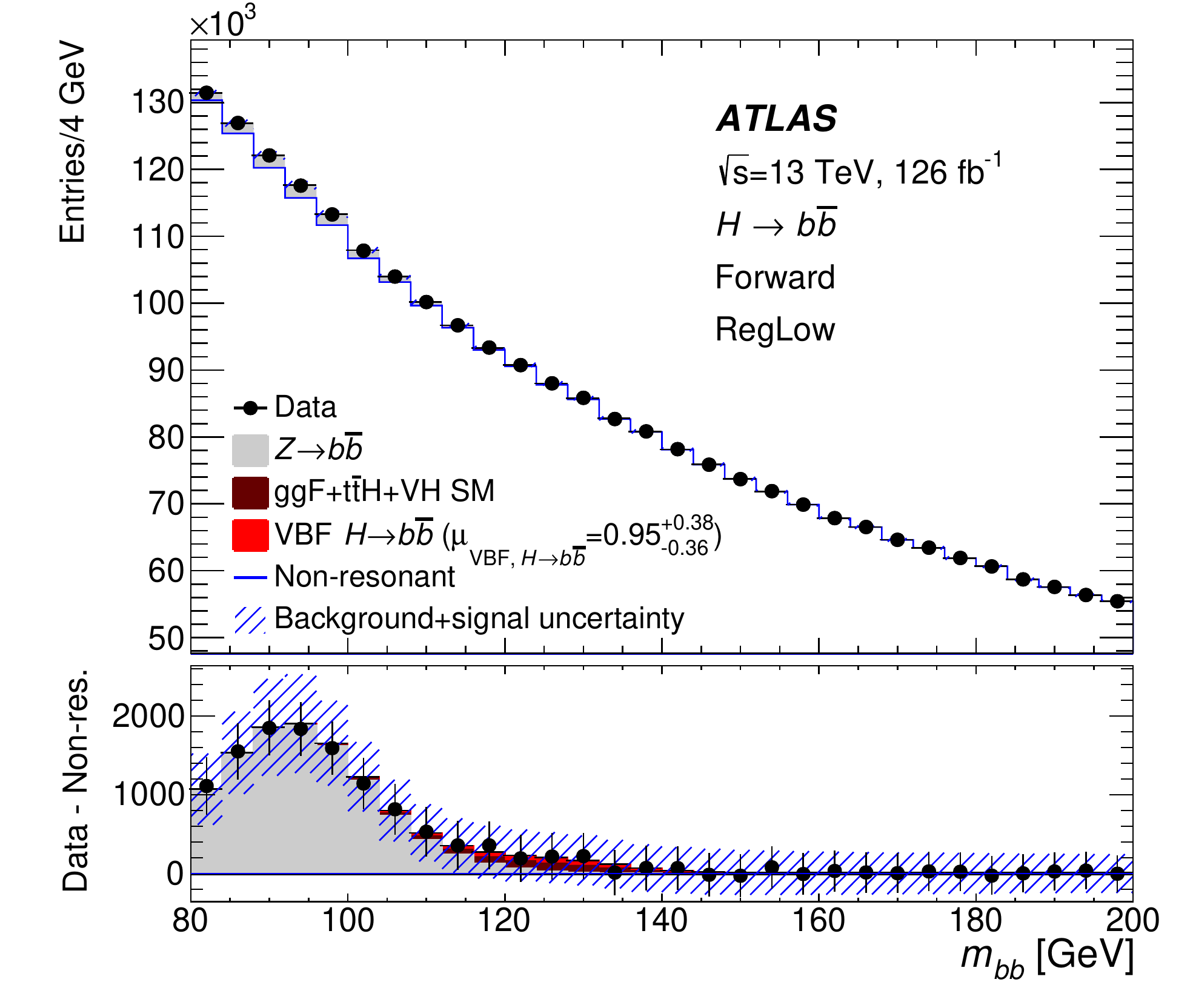}}
\caption{Fit results for the \twocentral{} channel.  Top left is SR1, top right is SR2, middle left is SR3, middle right is SR4, bottom is RegLow.  The data are the black points, the blue  line is the non-resonant background shape (continuum), grey is the resonant background ($Z\rightarrow b\bar{b}$), red is the fitted VBF Higgs boson signal (VBF $H\rightarrow b\bar{b}$), and maroon are events attributed to the other Higgs boson production mechanisms (ggF $+ \ttH + \VH$).  The bottom panel of each plot shows the difference between the data and the non-resonant background. The  hashed band shows the fitted background and signal uncertainties. For RegLow, the dominant background uncertainty is the statistical uncertainty of the shape of the background template. This uncertainty is mainly driven by the statistical uncertainty of data in RegLow, such that the background uncertainty is almost fully correlated with the data uncertainty in RegLow.}
\label{fig:unblind_1for2cen_pol_full}
\end{figure}
 
\begin{figure}[t!]\centering
\subfloat[]{\includegraphics[width=0.4\textwidth]{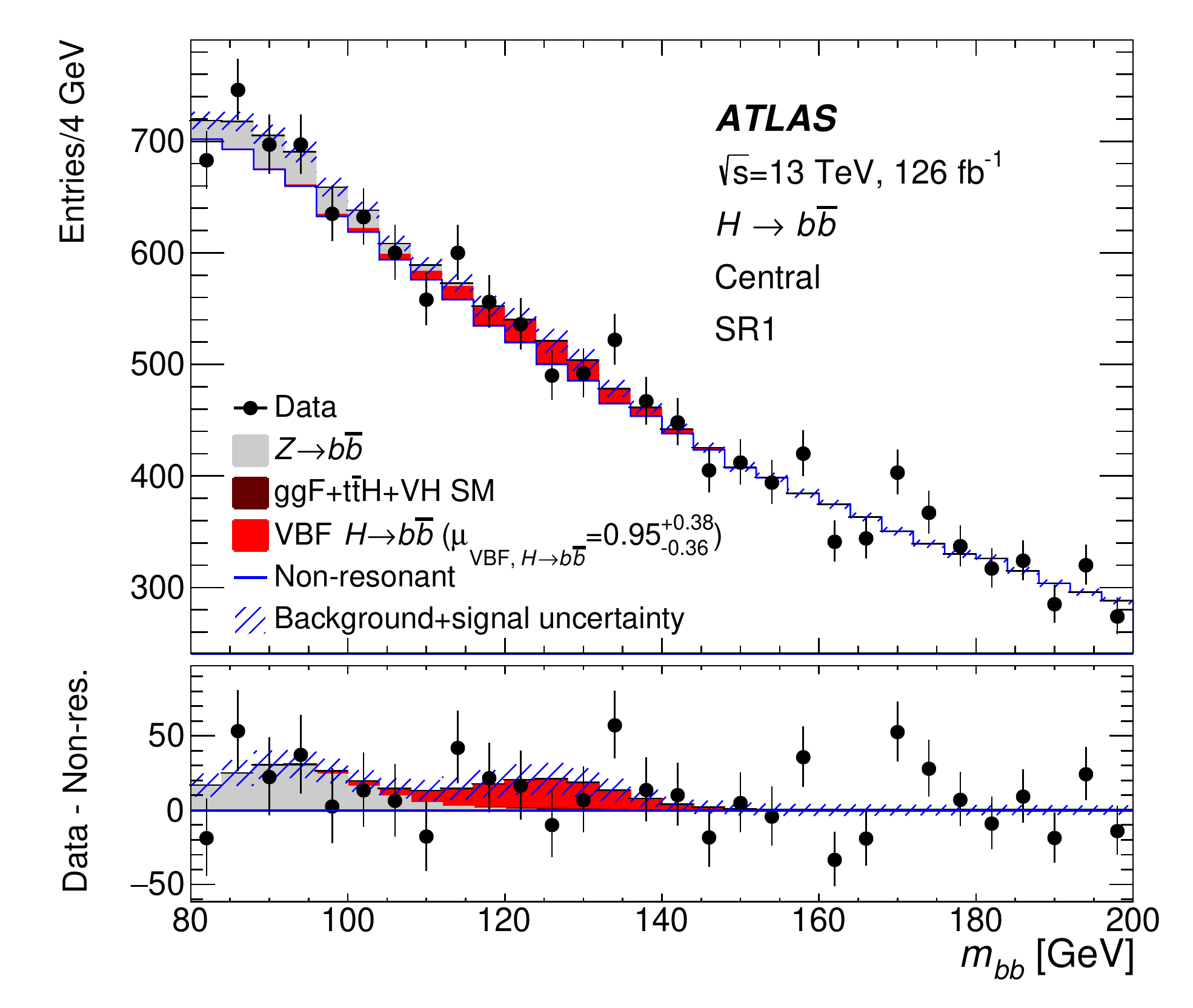}}
\subfloat[]{\includegraphics[width=0.4\textwidth]{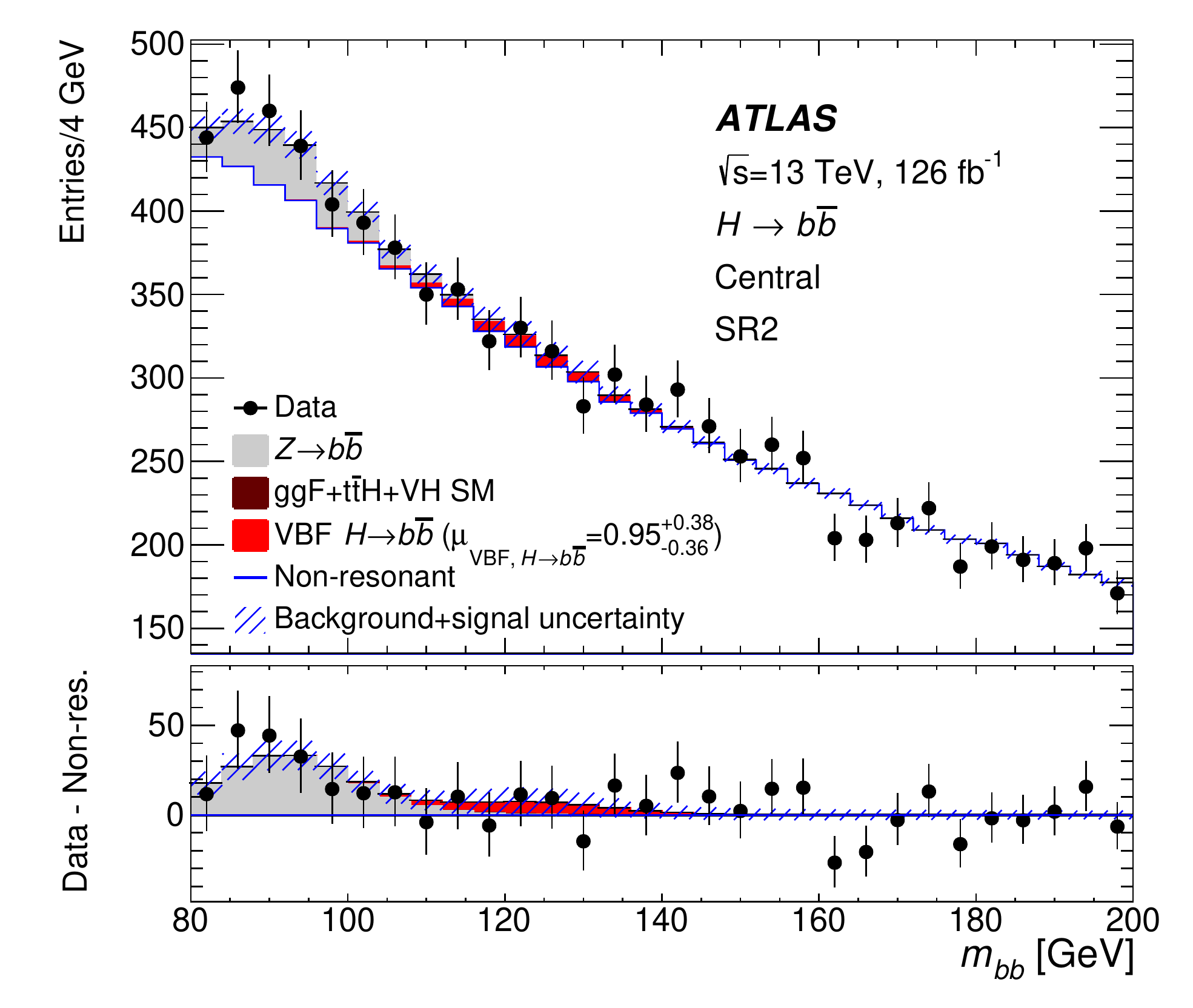}}\\
\subfloat[]{\includegraphics[width=0.4\textwidth]{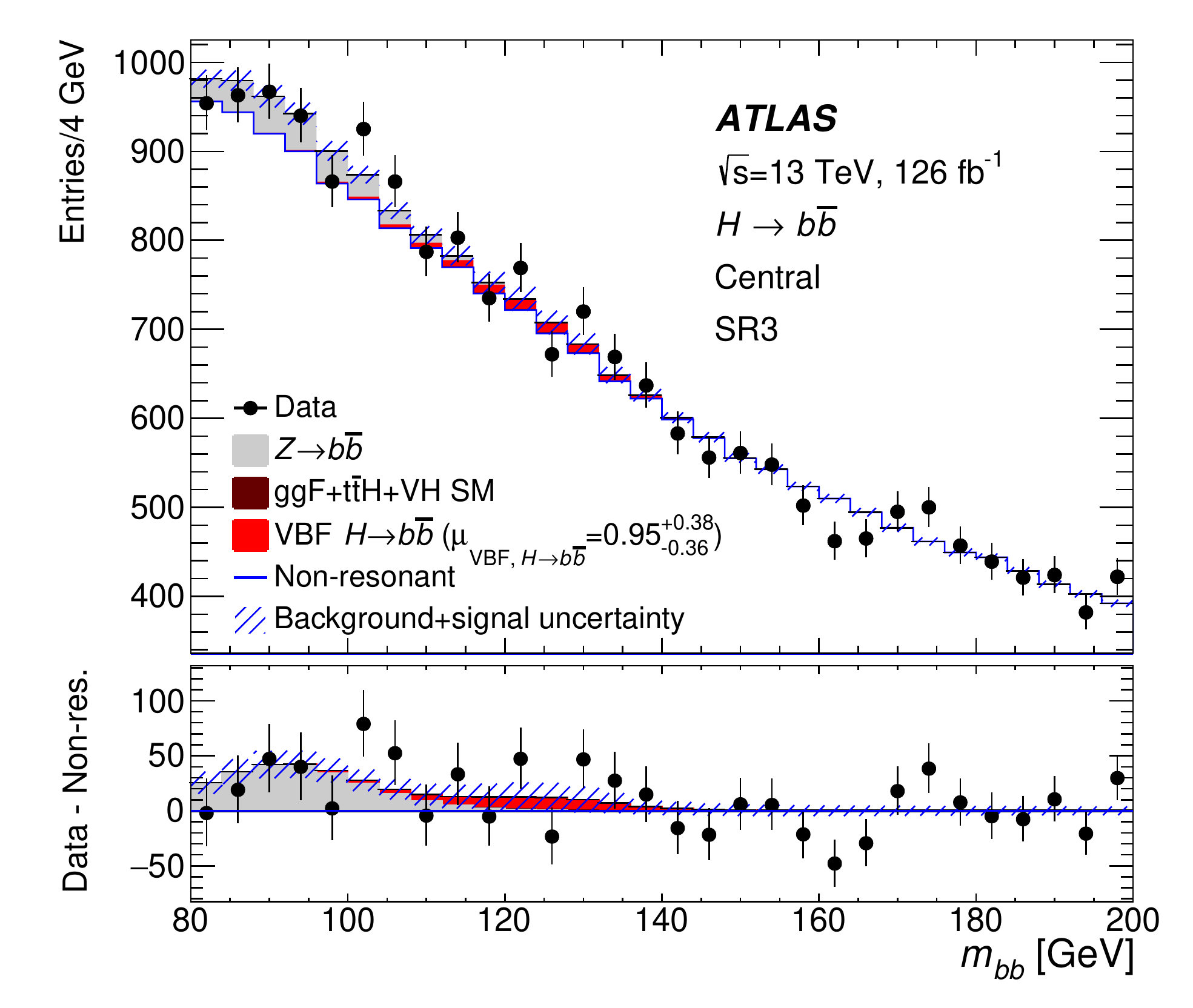}}
\subfloat[]{\includegraphics[width=0.4\textwidth]{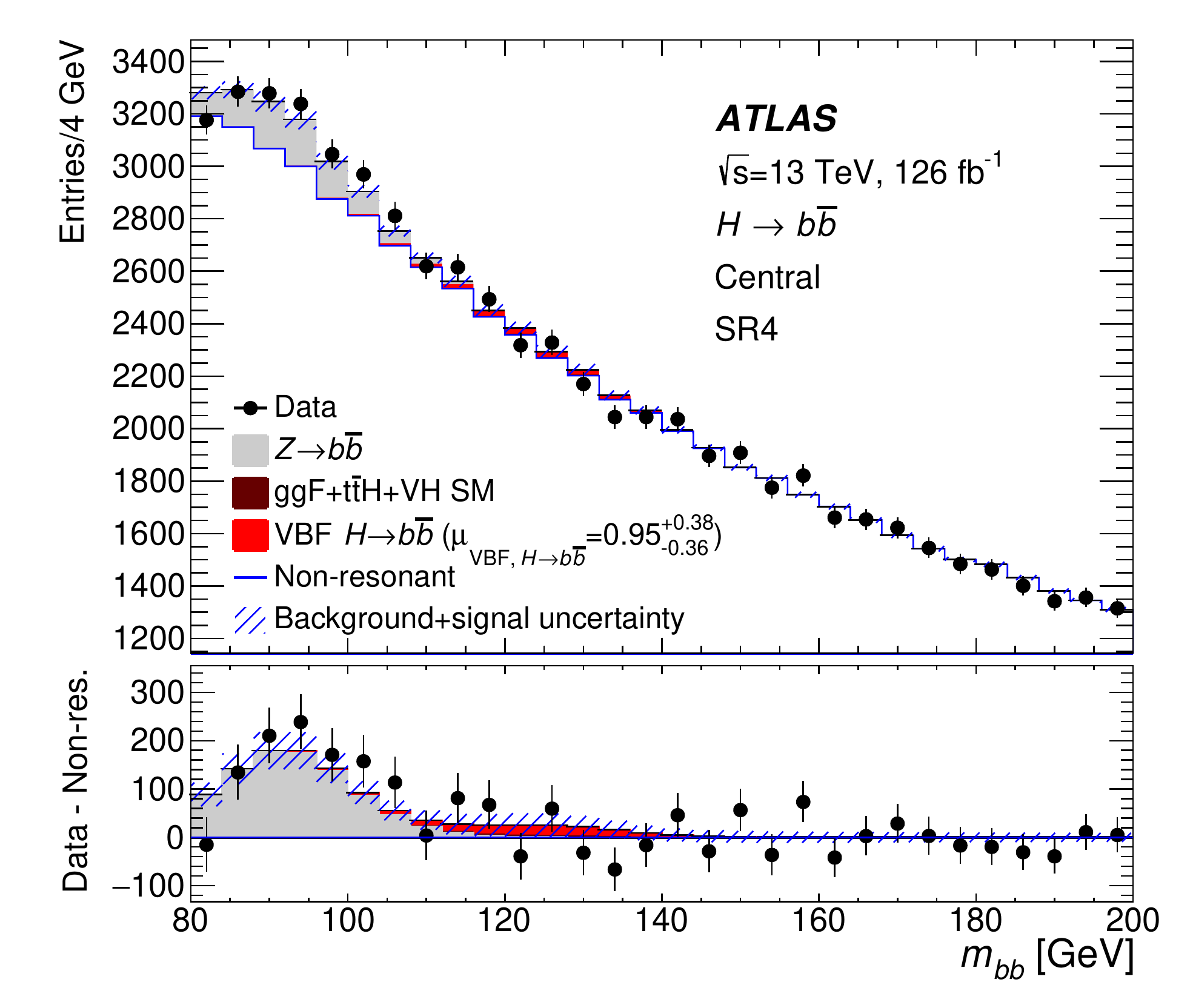}}\\
\subfloat[]{\includegraphics[width=0.4\textwidth]{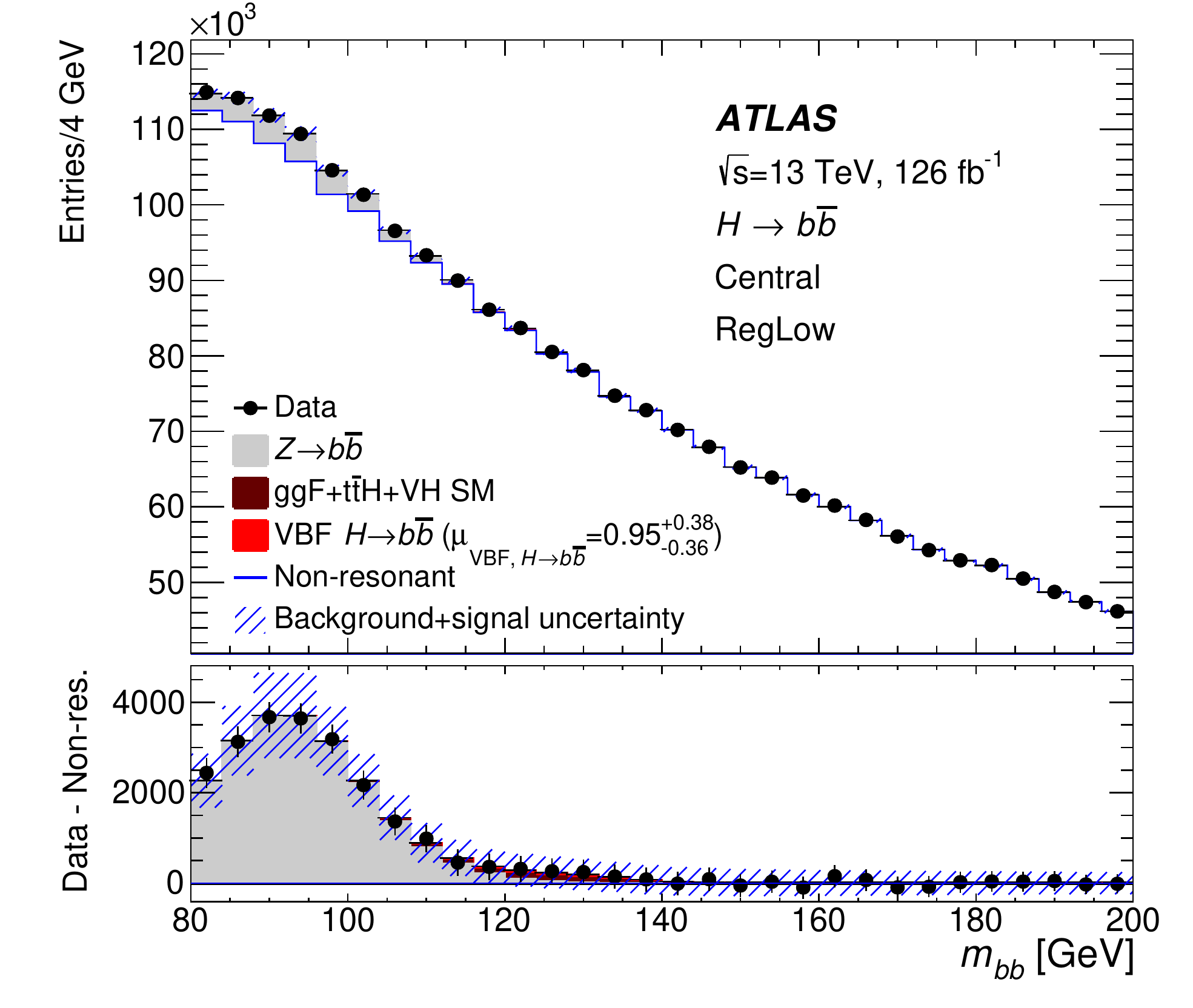}}
\caption{Fit results for the \threecentral{} channel.  Top left is SR1, top right is SR2, middle left is SR3, middle right is SR4, bottom is RegLow.  The data are the black points, the blue  line is the non-resonant background shape (continuum), grey is the resonant background ($Z\rightarrow b\bar{b}$), red is the fitted VBF Higgs boson signal  (VBF $H\rightarrow b\bar{b}$), and maroon are events attributed to the other Higgs boson production mechanisms (ggF $+ \ttH + \VH$).  The bottom panel of each plot shows the difference between the data and the non-resonant background. The hashed band shows the fitted background and signal uncertainties. For RegLow, the dominant background uncertainty is the statistical uncertainty of the shape of the background template. This uncertainty is mainly driven by the statistical uncertainty of data in RegLow, such that the background uncertainty is almost fully correlated with the data uncertainty in RegLow.}
\label{fig:unblind_3cen_pol_full}
\end{figure}
 
\begin{figure}[t!]\centering\
\includegraphics[width=0.8\textwidth]{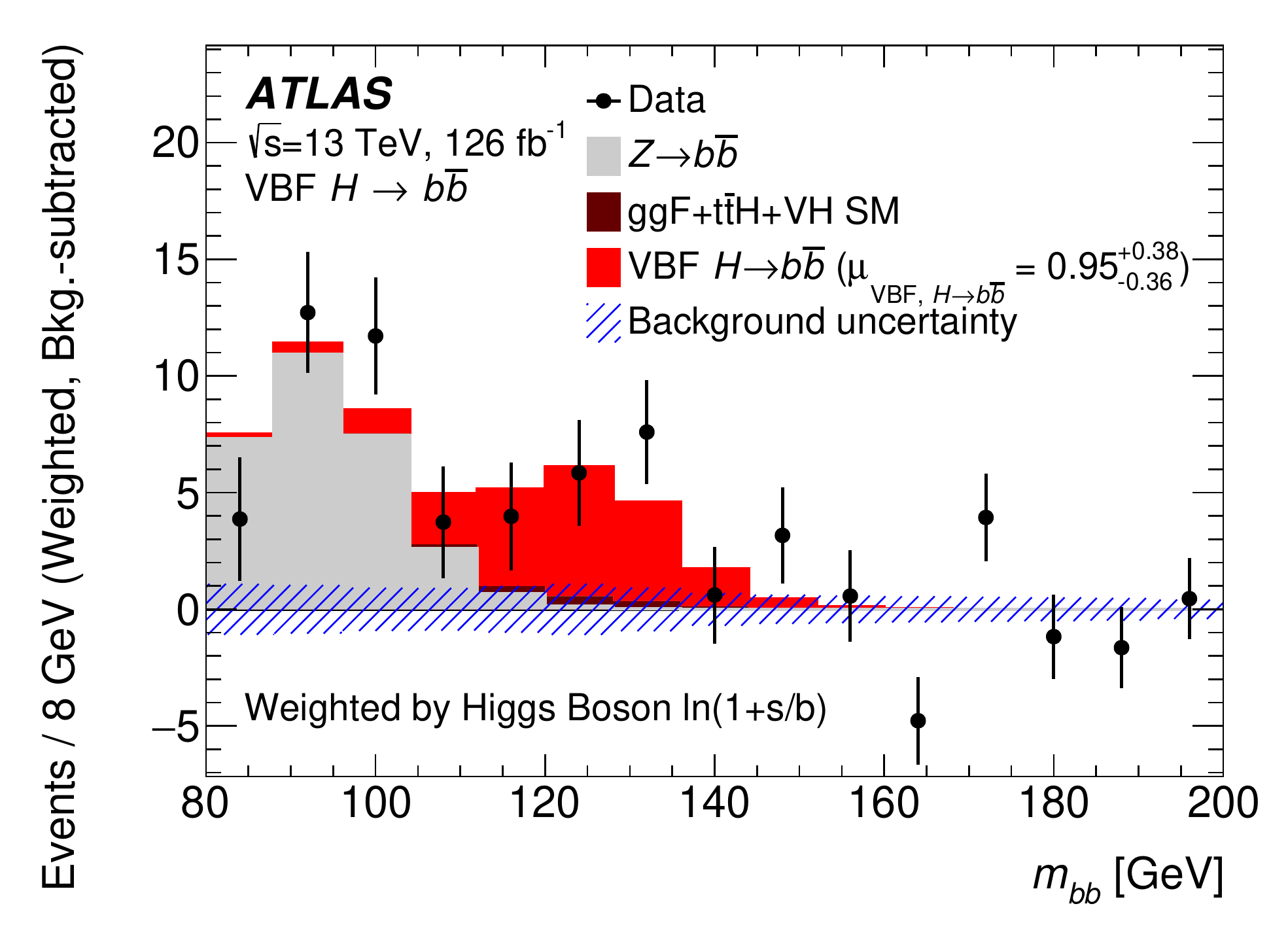}
\caption{The \mbb\ distribution after subtraction of non-resonant background, with signal regions 1 through 4 for both channels weighted by $\ln(1+s/b)$ within the range $112<\mbb<136~\GeV$ and summed. This range corresponds to roughly 68\% of the total Higgs boson signal template width. The value of $s$ is calculated from the observed Higgs boson signal (including all \Hbb production processes) within this range for each region, and $b$ is calculated from the post-fit background yields in each region.  The data are the black points, the  $Z$ contribution is in grey, the fitted VBF Higgs boson signal is in red, and maroon are events attributed to the other Higgs boson production mechanisms (ggF $+ \ttH + \VH$). The hashed band shows the fitted background uncertainties.} 
\label{fig:unblind_sOverBweight}
\end{figure}

\sloppy The observed $\muVBFHbb = \obsSSVBFSplit$ agrees well with the expectation, \expSSVBFSplit.  The significance of the measurement is \obsSigVBF$\sigma$ (\expSigVBF$\sigma$ expected).  The measurement of inclusive production yields similar results, with an observed \muH of \obsSSIncSplit compared to \expSSIncSplit expected, for a significance of \obsSigInc$\sigma$ (\expSigInc$\sigma$ expected).  Table~\ref{tab:higgs_significances_limits} summarizes the expected and observed significances and signal-strength parameters for both \muVBFHbb and \muH.  Table~\ref{tab:uncertainty_breakdown} summarizes the uncertainties in the \muVBFHbb measurement.  The measurement is statistically limited, with the largest uncertainties coming from the data statistics and the non-resonant background bias uncertainty.  The experimental systematic uncertainties, which are about a third as large as the statistical uncertainties, are dominated by the trigger and jet energy scale and resolution uncertainties.  Taken together, the theory uncertainties are comparable to the largest individual experimental uncertainties, as is the uncertainty due to the embedded $Z$ sample constraint. Due to the requirement on \pTbb, the measurement is primarily sensitive to higher-\pT Higgs boson production. The Higgs boson cross-section at high \pT is sensitive to many BSM models,
motivating the definition of dedicated high-\pT (true Higgs boson $\pT > 200\,\GeV$) regions in the simplified template cross-section framework~\cite{Berger:2019wnu}. The observed signal strength of inclusive Higgs boson production with a true $\pT > 200$~\GeV{} is \obsSSHPtSplit, for a significance of \obsSigHPt$\sigma$, compared with \expSigHPt$\sigma$ expected.  For this measurement, the contribution from events with true $\pT <200$~\GeV{} is fixed to its Standard Model expectation.  Results of two-dimensional likelihood scans of the signal strength for Higgs boson production with true boson \pT $> 200\,\GeV$ versus the signal strength for Higgs boson production with true boson $\pT < 200\,\GeV$ are shown in the Appendix, where the relative constraining power can be seen.  
Results of two-dimensional likelihood scans of the signal strength for Higgs boson production in the VBF production mode  versus the signal strength for Higgs boson production in the ggF production mode are also shown in the Appendix.   It can be seen that the measurement has little power to constrain ggF production separately from VBF production.
 
\FloatBarrier
 
\begin{table}[hbtp]
\small
\caption{Expected and observed significances and signal strengths for Higgs boson production, for  VBF production,  inclusive production, and production with a Higgs boson true \pT of greater than 200 \GeV, relative to the Standard Model prediction. }
\label{tab:higgs_significances_limits}
\centering
\def\arraystretch{1.4}
\begin{tabular}{l|r|r|r}
\hline
\hline
Results & VBF  Production &  Inclusive Production  & $\pT > 200$~\GeV{}\\
\hline
Expected significance   & $\expSigVBF \sigma$ & $\expSigInc \sigma$ & $\expSigHPt \sigma$ \\
Observed significance & $\obsSigVBF \sigma$ & $\obsSigInc \sigma$ & $\obsSigHPt \sigma$  \\
Expected signal strength  & $\expSSVBF$ & $\expSSInc $ & $\expSSHPt$  \\
Observed signal strength  & $\obsSSVBF$  & $\obsSSInc$ & $\obsSSHPt$ \\
\hline
\hline
\end{tabular}
\end{table}
 
\begin{table}[hbtp]
\small
\caption{Breakdown of the observed uncertainties on the inclusive production signal strength measurement (\muVBFHbb).}
\label{tab:uncertainty_breakdown}
\centering
\def\arraystretch{1.4}
\begin{tabular}{l|c}
\hline
\hline
Uncertainty & $\sigma(\muVBFHbb)$ \\
\hline
Statistics & $\pm0.32$ \\
NR Background Bias & $\pm0.15$  \\
Embedded $Z$ & $\pm0.05$  \\
\hline
Experimental & ${+0.10}/{-0.06}$  \\
~~~~Trigger & $+0.07/-0.03$  \\
~~~~Jet & $+0.07/-0.04$  \\
~~~~Flavor Tagging & $+0.02/-0.01$  \\
~~~~Other & $+0.03/-0.02$  \\
Signal Theory & $+0.06/-0.03$  \\
\hline
\hline
\end{tabular}
\end{table}

\FloatBarrier

The measured $\muVBF$ corresponds to a cross-section $\sigma_{\text{VBF}}\times \Brbb=\obsXsecBrSplit$~pb. Assuming a branching fraction $\Brbb=0.5809$~\cite{HiggsWG}, this corresponds to an inclusive VBF cross-section $\sigma_{\text{VBF}}=\obsXsecIncSplit$~pb.  
The predicted inclusive VBF cross-section calculated by the LHC Higgs Working Group at NNLO in QCD with proVBFH~\cite{ProVBFH}, including corrections at NLO from electroweak and photon processes determined with HAWK 2.0, is $3.77 \pm 0.08$~pb. A fiducial cross-section is determined by requiring $|y_{H}|<2.5$ and is measured to be  $ \sigma_{\text{VBF-fid}}= \obsXsecFidSplit$~fb.

These results are combined with the photon analysis~\cite{VBFgamma} described at the end of Section~\ref{sec:intro}.  The photon analysis yields a signal strength of $1.3\pm1.0$ with a significance of 1.3$\sigma$, compared with an expectation of 1.0$\sigma$ for both inclusive and VBF production. The two analyses are orthogonal, as the measurement described in this paper vetoes events with photons. They are combined in a joint likelihood fit, sharing the Higgs boson signal strength (\muVBFHbb or \muH) as a common parameter of interest.  All experimental systematic uncertainties are correlated with the exception of the jet energy resolution and offline flavour-tagging uncertainties, as the two analyses use different jet definitions.  The jet energy scale is partially correlated because both jet definitions use similar datasets to derive the uncertainties. Theoretical uncertainties are correlated except for the parton shower uncertainties, which are determined using a variation of the \HERWIG HardScale parameter in the photon analysis, and a comparison between \PYTHIA and \HERWIG in this analysis.   The combined measured signal strength for VBF production is  \obsSSVBFCombSplit, corresponding to a significance of \obsSigVBFComb$\sigma$ (\expSigVBFComb$\sigma$ expected).   This corresponds to an observed $\sigma_{\text{VBF}}\times \Brbb=\obsXsecBrCombSplit$~pb.   When fitting for inclusive Higgs boson production the observed signal strength is \obsSSIncCombSplit, corresponding to an observed (expected) significance of \obsSigIncComb$\sigma$ (\expSigIncComb$\sigma$).  The combination significances and signal strengths are summarized in Table~\ref{tab:higgs_significances_limits_comb}.
 
\begin{table}[hbtp]
\small
\caption{Expected and observed results for the Higgs boson production rate, for both inclusive production and VBF production only, relative to the Standard Model prediction for the combined analyses.}
\vspace{0.2cm}
\label{tab:higgs_significances_limits_comb}
\centering
\def\arraystretch{1.4}
\begin{tabular}{l|r|r}
\hline
\hline
Results  & VBF  Production & Inclusive Production \\
\hline
Expected significance  & $\expSigVBFComb \sigma$ & $\expSigIncComb \sigma$ \\
Observed significance & $\obsSigVBFComb \sigma$   & $\obsSigIncComb \sigma$ \\
Expected signal strength   & $\expSSVBFComb$ & $\expSSIncComb $ \\
Observed signal strength   & $\obsSSVBFComb$ & $\obsSSIncComb$ \\
\hline
\hline
\end{tabular}
\end{table}


\FloatBarrier

\section{Conclusion}
This paper presents a measurement of Standard Model VBF Higgs boson production in the $b\bar{b}$ decay mode using 126~\ifb\ of 13~\TeV\ proton--proton data collected with the ATLAS detector at the LHC.  Significant improvements relative to the previous analysis using 31~\ifb\, lead to a measurement of VBF Higgs boson production with a significance of \obsSigVBF $\sigma$ relative to the background-only hypothesis, with an observed signal strength of \obsSSVBFSplit compared to an expected value of \expSSVBFSplit.   The improvements include the usage of an ANN to decorrelate \mbb{} from the event classifier, boosting the statistical power of the fit by allowing the same background shape to be used in all signal regions.  Additionally, the usage of embedding to estimate the $Z$ boson contribution directly from data  results in significantly less uncertainty on the $Z$ boson contribution to the fit. The event selection and classifier provide sensitivity primarily to VBF production at high Higgs boson~\pT.  The measured cross-section is $\sigma_{\text{VBF}}\times \Brbb=\obsXsecBrSplit$~fb, and the  inclusive VBF cross-section $\sigma_{\text{VBF}}=\obsXsecIncSplit$~fb. Requiring $|y_{H}|<2.5$, this corresponds to a fiducial cross-section, $ \sigma_{\text{VBF-fid}} = \obsXsecFidSplit$~fb.  These results are combined with a complementary measurement of VBF $H\rightarrow b\bar{b}$ production in association with a photon, leading to a \obsSigVBFComb$\sigma$ measurement of VBF $H\rightarrow b\bar{b}$ production.

 
\FloatBarrier
\section*{Acknowledgements}
 

We thank CERN for the very successful operation of the LHC, as well as the
support staff from our institutions without whom ATLAS could not be
operated efficiently.
 
We acknowledge the support of ANPCyT, Argentina; YerPhI, Armenia; ARC, Australia; BMWFW and FWF, Austria; ANAS, Azerbaijan; SSTC, Belarus; CNPq and FAPESP, Brazil; NSERC, NRC and CFI, Canada; CERN; ANID, Chile; CAS, MOST and NSFC, China; COLCIENCIAS, Colombia; MSMT CR, MPO CR and VSC CR, Czech Republic; DNRF and DNSRC, Denmark; IN2P3-CNRS and CEA-DRF/IRFU, France; SRNSFG, Georgia; BMBF, HGF and MPG, Germany; GSRT, Greece; RGC and Hong Kong SAR, China; ISF and Benoziyo Center, Israel; INFN, Italy; MEXT and JSPS, Japan; CNRST, Morocco; NWO, Netherlands; RCN, Norway; MNiSW and NCN, Poland; FCT, Portugal; MNE/IFA, Romania; JINR; MES of Russia and NRC KI, Russian Federation; MESTD, Serbia; MSSR, Slovakia; ARRS and MIZ\v{S}, Slovenia; DST/NRF, South Africa; MICINN, Spain; SRC and Wallenberg Foundation, Sweden; SERI, SNSF and Cantons of Bern and Geneva, Switzerland; MOST, Taiwan; TAEK, Turkey; STFC, United Kingdom; DOE and NSF, United States of America. In addition, individual groups and members have received support from BCKDF, CANARIE, Compute Canada, CRC and IVADO, Canada; Beijing Municipal Science \& Technology Commission, China; COST, ERC, ERDF, Horizon 2020 and Marie Sk{\l}odowska-Curie Actions, European Union; Investissements d'Avenir Labex, Investissements d'Avenir Idex and ANR, France; DFG and AvH Foundation, Germany; Herakleitos, Thales and Aristeia programmes co-financed by EU-ESF and the Greek NSRF, Greece; BSF-NSF and GIF, Israel; La Caixa Banking Foundation, CERCA Programme Generalitat de Catalunya and PROMETEO and GenT Programmes Generalitat Valenciana, Spain; G\"{o}ran Gustafssons Stiftelse, Sweden; The Royal Society and Leverhulme Trust, United Kingdom.
 
The crucial computing support from all WLCG partners is acknowledged gratefully, in particular from CERN, the ATLAS Tier-1 facilities at TRIUMF (Canada), NDGF (Denmark, Norway, Sweden), CC-IN2P3 (France), KIT/GridKA (Germany), INFN-CNAF (Italy), NL-T1 (Netherlands), PIC (Spain), ASGC (Taiwan), RAL (UK) and BNL (USA), the Tier-2 facilities worldwide and large non-WLCG resource providers. Major contributors of computing resources are listed in Ref.~\cite{ATL-SOFT-PUB-2020-001}.
 

\clearpage
 
\appendix
\part*{Appendix}
\addcontentsline{toc}{part}{Appendix}
\FloatBarrier
This analysis is particularly sensitive to high Higgs boson \pT where true Higgs boson \pT, \pTH,  is greater than 200\,\GeV, but is unable to separately constrain the cross-sections at high and low \pT.
In the forward channel, 54\% of events have both $\pTbb$ and $\pTH >$  200 \GeV, 34\% have both $\pTbb$ and $\pTH <$ 200\, \GeV \, and 8 (4)\% have $\pTbb < (>)$ 200\,\GeV \, and  $\pTH > (<)$ 200\, \GeV. In the central channel, 79\% of events have both $\pTbb$ and $\pTH >$ 200\, \GeV, 13\% have both $\pTbb$ and $\pTH <$ 200 \,\GeV, and 7 (2)\% have $\pTbb < (>)$ 200\,\GeV \, and  $\pTH > (<)$ 200\, \GeV.  The two-dimensional likelihood scan of high- and low-\pT Higgs boson production is shown in Figure~\ref{fig:scan_ptsplit},
where the relative size of the ranges for each \pT bin gives a measure of the relative sensitivity to high and low Higgs boson \pT.  Because the total VBF contribution is rather well-constrained, the production at high and low \pT is anti-correlated.
 
This analysis does not strongly constrain the ggF Higgs boson production cross-section,
as the ggF kinematics are similar to the kinematics of the SM background, resulting in a similar ANN score distribution as the background.
The VBF production cross-section cannot be measured precisely if the ggF cross-section is unconstrained,
particularly due to large uncertainties for ggF events with at least two extra jets and high \mJJ.
The two-dimensional likelihood scan of VBF and ggF production signal strengths is shown in Figure~\ref{fig:scan_ggfvbf}.
A large range in the ggF signal strength is observable in the figure,
and the impact of the lack of ggF constraint on the VBF production signal strength is only
found to be large when the ggF production signal strength is far from the SM expectation.
 
\begin{figure}[h]\centering
\includegraphics[width=0.8\textwidth]{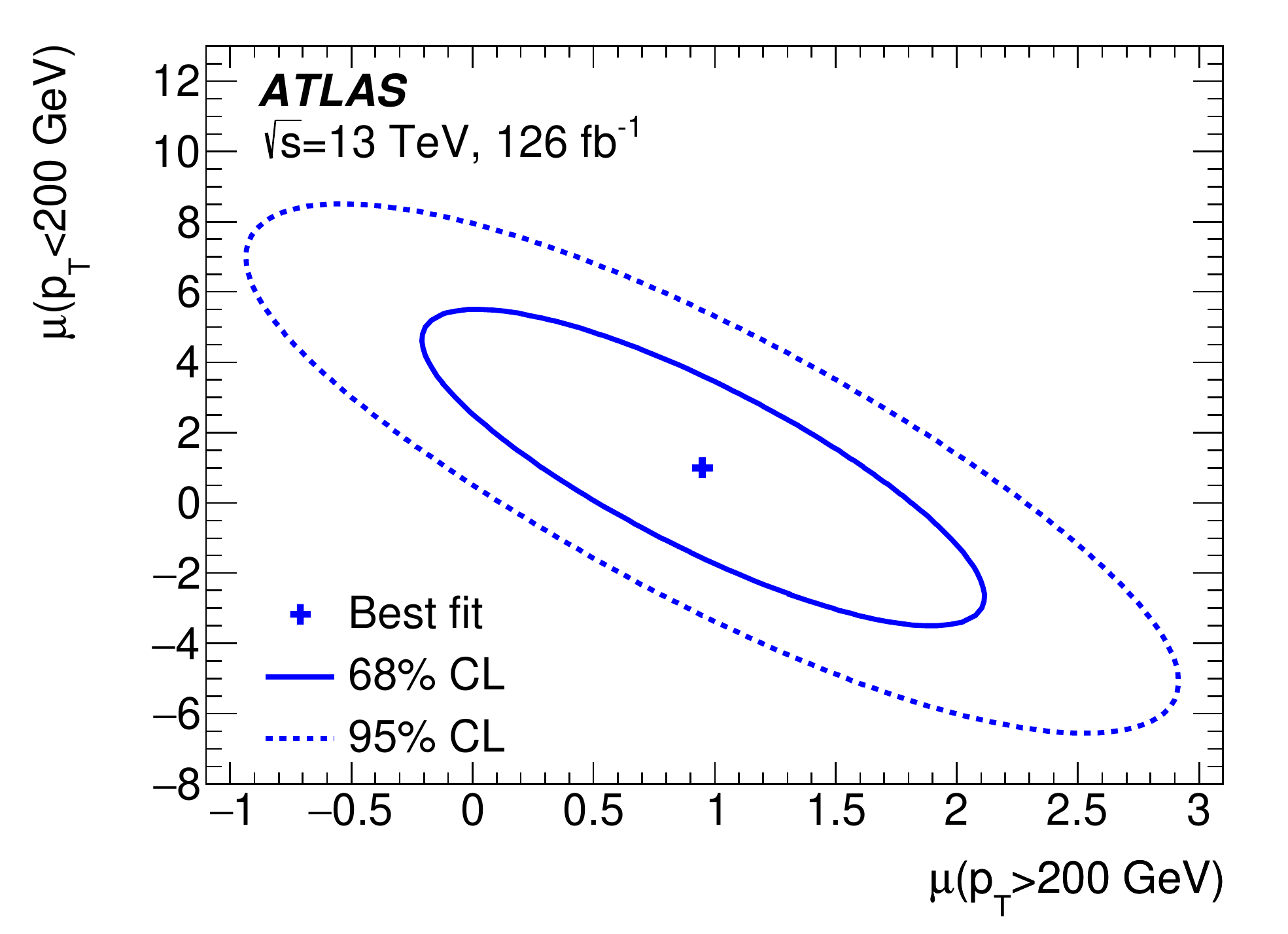}
\caption{Likelihood scan as a function of the signal strength for Higgs boson production with true boson $\pT > 200\,\GeV$ versus the signal strength for Higgs boson production with true boson $\pT < 200\,\GeV$.  Shown is the best-fit value (blue plus sign), and the 68\% (solid blue) and 95\% (dashed blue) confidence level (CL) contours.}
\label{fig:scan_ptsplit}
\end{figure}
 
\begin{figure}[h]\centering
\includegraphics[width=0.8\textwidth]{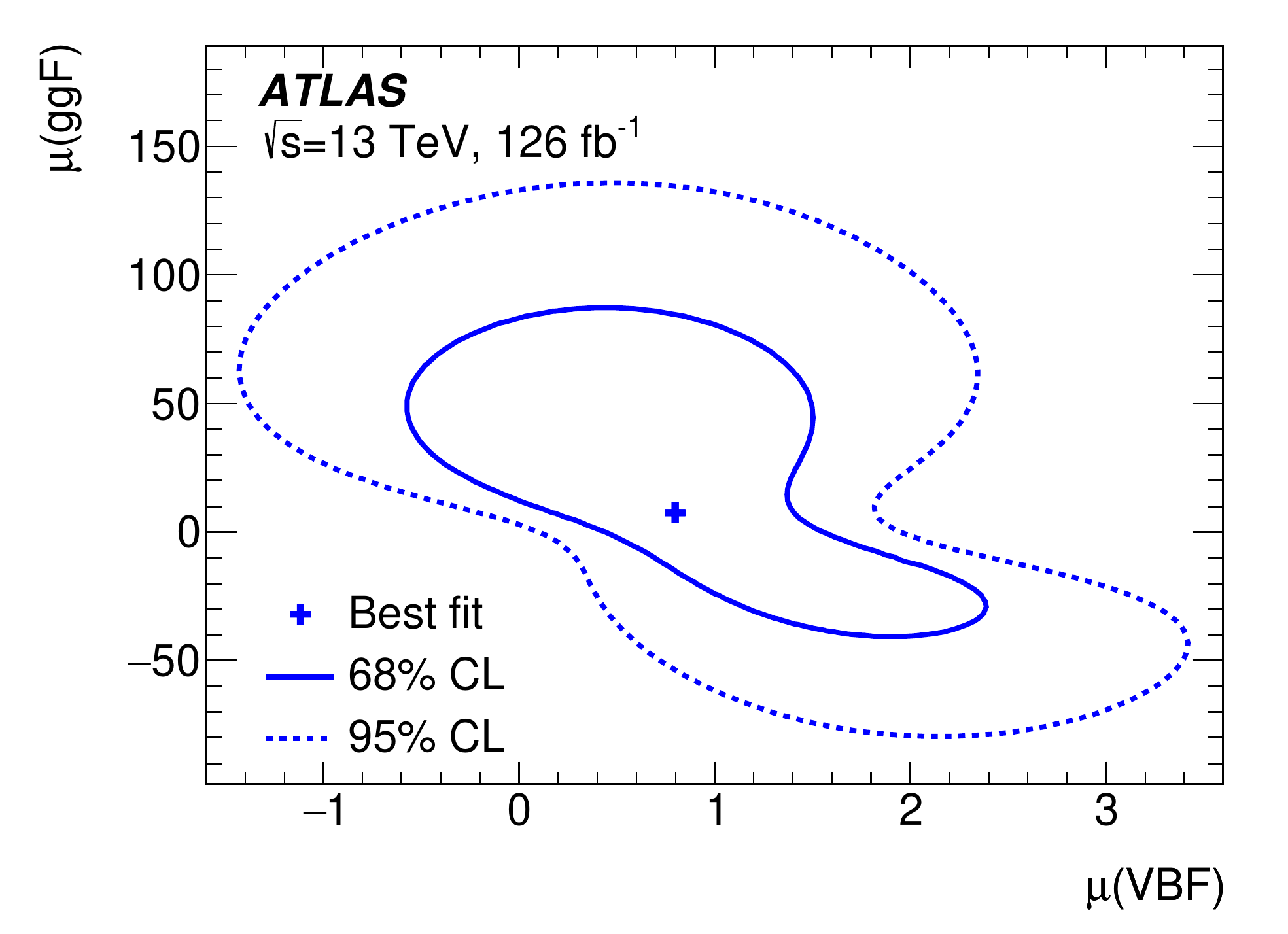}
\caption{Likelihood scan as a function of the signal strength for VBF Higgs boson production versus the signal strength for ggF Higgs boson production.  Shown is the best-fit value (blue plus sign), and the 68\% (solid blue) and 95\% (dashed blue) confidence level (CL) contours. The asymmetric shape arises from the fit's inability to separately constrain the subset of ggF events which have a VBF-like topology.}
\label{fig:scan_ggfvbf}
\end{figure}
 
\FloatBarrier

\printbibliography
\FloatBarrier
\clearpage

\clearpage 
 
\begin{flushleft}
\hypersetup{urlcolor=black}
{\Large The ATLAS Collaboration}

\bigskip

\AtlasOrcid[0000-0002-6665-4934]{G.~Aad}$^\textrm{\scriptsize 101}$,    
\AtlasOrcid[0000-0002-5888-2734]{B.~Abbott}$^\textrm{\scriptsize 127}$,    
\AtlasOrcid[0000-0002-7248-3203]{D.C.~Abbott}$^\textrm{\scriptsize 102}$,    
\AtlasOrcid[0000-0002-2788-3822]{A.~Abed~Abud}$^\textrm{\scriptsize 36}$,    
\AtlasOrcid[0000-0002-1002-1652]{K.~Abeling}$^\textrm{\scriptsize 53}$,    
\AtlasOrcid[0000-0002-2987-4006]{D.K.~Abhayasinghe}$^\textrm{\scriptsize 93}$,    
\AtlasOrcid[0000-0002-8496-9294]{S.H.~Abidi}$^\textrm{\scriptsize 29}$,    
\AtlasOrcid[0000-0002-8279-9324]{O.S.~AbouZeid}$^\textrm{\scriptsize 40}$,    
\AtlasOrcid{N.L.~Abraham}$^\textrm{\scriptsize 155}$,    
\AtlasOrcid[0000-0001-5329-6640]{H.~Abramowicz}$^\textrm{\scriptsize 160}$
% ...
\bigskip
\\

% ...
$^{29}$Physics Department, Brookhaven National Laboratory, Upton NY; United States of America.\\
% ...
$^{36}$CERN, Geneva; Switzerland.\\
% ...
$^{40}$Niels Bohr Institute, University of Copenhagen, Copenhagen; Denmark.\\
% ...
$^{53}$II. Physikalisches Institut, Georg-August-Universit\"{a}t G\"ottingen, G\"ottingen; Germany.\\
% ...
$^{93}$Department of Physics, Royal Holloway University of London, Egham; United Kingdom.\\
% ...
$^{101}$CPPM, Aix-Marseille Universit\'e, CNRS/IN2P3, Marseille; France.\\
$^{102}$Department of Physics, University of Massachusetts, Amherst MA; United States of America.\\
% ...
$^{127}$Homer L. Dodge Department of Physics and Astronomy, University of Oklahoma, Norman OK; United States of America.\\
% ...
$^{155}$Department of Physics and Astronomy, University of Sussex, Brighton; United Kingdom.\\
% ...
$^{160}$Raymond and Beverly Sackler School of Physics and Astronomy, Tel Aviv University, Tel Aviv; Israel.\\
% ...

\end{flushleft}


\end{document}